\documentclass[runningheads]{llncs}
\usepackage{times}
\usepackage{amsmath,amsfonts,amssymb, mathbbol}
\usepackage{extarrows}
\usepackage{stmaryrd}
\usepackage{comment}
\usepackage[font=small]{caption}
\usepackage[labelformat=simple]{subcaption}

\usepackage{wrapfig}

\usepackage[inline]{enumitem}
\usepackage[ruled,vlined,linesnumbered]{algorithm2e}
\SetAlFnt{\scriptsize} 
\usepackage[usenames, table, svgnames, dvipsnames]{xcolor}
\usepackage{url}
\usepackage[title,page]{appendix}
\usepackage{algcompatible}
\usepackage{tabularx}

\usepackage{booktabs}
\usepackage{array}
\usepackage{multirow}

\usepackage{tikz}
\usetikzlibrary{arrows,shapes,snakes,automata,backgrounds,petri,positioning, calc}
\usepackage{graphicx}

\usepackage[hyperfootnotes=true,colorlinks,linkcolor=RedViolet,anchorcolor=black,citecolor=blue,urlcolor=blue]{hyperref} 

\newtheorem{constraint}{Constraint}

\newcommand{\oomit}[1]{}

\newcolumntype{Y}{>{\centering\arraybackslash}X}

\usepackage{todonotes}

\usepackage{xspace}

\newcommand{\ssr}{S\cup S_+ \cup R}
\newcommand{\bv}{\mathbb{b}}
\newcommand{\qv}{\mathbb{q}}

\newcommand\xqed[1]{%
  \leavevmode\unskip\penalty9999 \hbox{}\nobreak\hfill
  \quad\hbox{#1}}
\newcommand\demo{\xqed{$\lhd$}}

\makeatletter

\def\orcidID#1{\smash{\href{http://orcid.org/#1}{\protect\raisebox{-1.25pt}{\protect\includegraphics{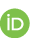}}}}}
\makeatother

\begin{document}
\title{Active Learning of One-Clock Timed Automata using Constraint Solving\thanks{This work has been partially funded by NSFC under grant No.~62032024, and by DFG project 389792660-TRR 248.}}
%
%
\author{Runqing Xu\inst{1,2}\orcidID{0000-0003-1521-7379} \and
	Jie An\inst{3}\orcidID{0000-0001-9260-9697} \and
	Bohua Zhan\inst{1,2}\orcidID{0000-0001-5377-9351}}
%
%
\institute{
State Key Lab. of Computer Science, Institute of Software, CAS, Beijing, China \and
University of Chinese Academy of Sciences, Beijing, China\\
\email{\{xurq,bzhan\}@ios.ac.cn}
\and
Max Planck Institute for Software Systems, Kaiserslautern, Germany\\
\email{jiean@mpi-sws.org}
}
\maketitle              
\begin{abstract}
Active automata learning in the framework of Angluin's $L^*$ algorithm has been applied to learning many kinds of automata models. In applications to timed models such as timed automata, the main challenge is to determine guards on the clock value in transitions as well as which transitions reset the clock. In this paper, we introduce a new algorithm for active learning of deterministic one-clock timed automata and timed Mealy machines. The algorithm uses observation tables that do not commit to specific choices of reset, but instead rely on constraint solving to determine reset choices that satisfy readiness conditions. We evaluate our algorithm on randomly-generated examples as well as practical case studies, showing that it is applicable to larger models, and competitive with existing work for learning other forms of timed models.
	
\keywords{Active Learning \and Timed Automata \and Constraint Solving.}
\end{abstract}
\section{Introduction}\label{sec:introduction}

Within Angluin's $L^*$ framework~\cite{Angluin87}, active learning is a type of model inference to learn an unknown language by making queries to a teacher. There are two kinds of queries: membership queries and equivalence queries. For a membership query, the teacher answers whether the queried word is in the target language. Usually, the learner collects query results in an \emph{observation table}. When the observation table satisfies some readiness conditions, it can be transformed to a candidate automaton for an equivalence query. The teacher answers whether the candidate automaton recognizes the target language, and returns a counterexample if the answer is negative. 
In recent decades, the core algorithm has seen many technical improvements, has been extended to learn different kinds of models, and has been applied to many realistic settings. We refer to~\cite{SteffenHM11,Vaandrager17} for surveys.

For timed systems, timing constraints play a key role in the correctness of the system. In general, automata learning of timed systems require learning a timed model from either passive or active observations of the system, consisting of a collection of time-event sequences. The learned model should describe these timing behaviors correctly. Timed automata~\cite{Alur94}, extending DFAs with clock variables, is a popular formal model of timed systems. However, there are several obstacles to active learning of timed automata. Since the transitions of timed automata contain both timing constraints that test the values of clocks, and resets that update the clocks, we need to determine (1) the number of clocks, (2) the reset information, and (3) the timing constraints, none of which are directly observable from time-event sequences. Hence, existing work consider timed automata with different restrictions. 
Among them, An et al. introduced an active learning algorithm for deterministic one-clock timed automata (DOTAs)~\cite{AnCZZZ20}. They first suppose that the teacher can return reset information in the queries, then the assumption is dropped by allowing the algorithm to search through possible combinations of reset information. However, this search process results in an exponential blow up, limiting the scalability of the algorithm in practical applications. Vaandrager et al.~\cite{VaandragerB021} considered a different class of timed models called Mealy machines with one timer, and proposed a learning algorithm with polynomial complexity.

In this paper, we present a new active learning algorithm for deterministic one-clock timed automata. The main innovation of the algorithm is to maintain all available observations in a single observation table, without committing to a particular choice of resets. The readiness conditions of the observation table, such as closedness, consistency, etc, are encoded as formulas in terms of variables for reset information and location assignments. These constraints are then solved using SMT solvers to determine feasible assignments of resets and locations that make the observation table ready, and from which a candidate automaton can be constructed. The learning algorithm is guaranteed to terminate and return a correct automaton. While the theoretical worst-case complexity of the algorithm is still exponential, by leveraging the efficiency of SMT solvers, it is much more efficient than the algorithm in~\cite{AnCZZZ20} in practice. In order to apply the algorithm to learning real-time reactive systems, we extend it to timed Mealy machines, which can be considered as an extension of Mealy machines with one clock, or extension of deterministic one-clock timed automata to include inputs and outputs.

The algorithm is implemented and evaluated on a number of randomly generated models and four models from practical applications. The experimental results show that our algorithm is scalable to much larger models compared to~\cite{AnCZZZ20}. Additionally, our method successfully learns all four models from practical applications, with costs that is competitive against algorithms designed for other forms of models.

The organization of the paper is as follows. We give some background material in Section~\ref{sec:preliminaries}. The algorithm for learning DOTAs is described in Section~\ref{sec:learning_CS}, and its extension to timed Mealy machines in Section~\ref{sec:mealy}. We describe the implementation and experiments in Section~\ref{sec:experiment}, and finally conclude in Section~\ref{sec:conclusion}.

\paragraph*{Related work.} Active learning of timed systems has been studied on many kinds of models with different restrictions. In~\cite{GrinchteinJP06,Grinchtein10}, Grinchtein et al. proposed learning algorithms for event-recording automata (ERAs)~\cite{AlurFH99}, a kind of timed automata associating every action $a$ with a clock $x_a$ that records the length of time since the last occurrence of $a$.  
Henry et al. considered in~\cite{HenryJM20} reset-free ERAs, where some transitions may reset no clocks. 
An et al. introduced a learning algorithm for deterministic one-clock timed automata (DOTAs) in~\cite{AnCZZZ20}, but due to the brute-force search over choice of resets, the algorithm is limited to timed automata with a small number of locations.
For real-time automata (RTA)~\cite{Dima01}, efficient learning algorithms have been designed in both the deterministic~\cite{AnWZZZ21} and nondeterministic case~\cite{AnZZZ21}. Recently, Vaandrager et al. introduced a new kind of timed models named Mealy machine with one timer (MM1T), and proposed an efficient active learning algorithm for such models~\cite{VaandragerB021}. It extends Mealy machine with a single timer which can be set to an integer value at transitions.

Passive learning has also been investigated for learning timed automata based on different methods~\cite{VerwerWW09,VerwerWW11,VerwerWW12,TapplerALL19,AichernigPT20}. Constraint solving has been used extensively in passive learning. For example, Smetsers et al. used this technique in passive learning of DFAs, Mealy machines and register automata~\cite{SmetsersFV18}, by encoding the existence of an automaton with $n$ locations consistent with a set of observations in a logical formula. Recent work~\cite{TapplerAL22} also applied constraint solving to passive learning of timed automata. Compared to these works, we remain in the active learning setting, encoding constraints for the readiness of observation tables rather than consistency with a set of observations. Our work demonstrates that constraint solving can be fruitfully applied in active learning, determining not only location assignments but also other hidden parts of the model such as clock-reset information.

\section{Preliminaries}\label{sec:preliminaries}

In this section, we introduce several concepts of one-clock timed automata. Let $\mathbb{R}_{\geq 0}$ and $\mathbb{N}$ be the set of non-negative reals and natural numbers, respectively. The set of boolean values is denoted as $\mathbb{B}=\{\top,\bot\}$, where $\top$ stands for \emph{true} and $\bot$ for \emph{false}.

Let $c$ be the single clock variable, denote by $\Phi_c$ the set of clock constraints of the form $\phi::= \top \mid c\bowtie m \mid \phi \wedge \phi$, where $m\in\mathbb{N}$ and $\bowtie\ \in \{=,<,>,\le,\ge\}$. Since there is only one clock, a clock constraint can be represented as an integer-bounded interval whose endpoints are in $\mathbb{N}\cup\{\infty\}$. For example, $c\leq5 \wedge c > 4$ is represented as $(4,5]$, $c=6$ as $[6,6]$, and $\top$ as $[0,\infty)$. We will use inequality and interval representations interchangeably in this paper. Let the finite set of actions $\Sigma$ be fixed.

\begin{definition}[One-clock timed automata~\cite{AnCZZZ20}]\label{def:otas}
	A one-clock timed automaton (OTA) $\mathcal{A}$ is a 6-tuple $(\Sigma,Q,q_0,F,c,\Delta)$, where $\Sigma$ is a finite set of actions, called the \emph{alphabet}; $Q$ is a finite set of locations; $q_0\in Q$ is the initial location; $F\subseteq Q$ is a set of accepting locations; $c$ is the unique clock; and $\Delta\subseteq Q\times\Sigma\times\Phi_c\times\mathbb{B} \times Q$ is a finite set of transitions.
\end{definition}

A transition $\delta = (q,\sigma,\phi,b,q')$ allows a jump from  the \emph{source location} $q$ to the \emph{target location} $q'$ by performing the action $\sigma\in\Sigma$ if the \emph{guard} $\phi\in \Phi_c$ is satisfied. Meanwhile, clock $c$ is reset to zero if $b = \top$, and remains unchanged otherwise.
\emph{Clock valuation} is a function $\nu\colon c \rightarrow \mathbb{R}_{\ge 0}$ that assigns a non-negative real number to the clock. 
A \emph{state} of $\mathcal{A}$ is a pair $(q,\nu)$, where $q\in Q$ and $\nu$ is a clock valuation.

Given an OTA $\mathcal{A}$, with $\kappa$ being the maximum constant appearing in the guards, then the clock valuations can be divided into \emph{regions}, where each region is of the form $[n,n]$ for $n\le\kappa$, or $(n,n+1)$ for $n<\kappa$, or $(\kappa,\infty)$. This gives a partition of $\mathbb{R}_{\ge 0}$. For clock valuation $\nu$, we denote by $\llbracket \nu \rrbracket$ the region containing it. Regions are commonly used in algorithms for analyzing timed automata, making the state space essentially finite.

Given a \emph{timed word} $\omega=(\sigma_1,t_1)(\sigma_2,t_2)\cdots(\sigma_n,t_n)\in (\Sigma\times\mathbb{R}_{\geq 0})^*$, where $t_i$ represents the delay time between two actions $\sigma_{i-1}$ and $\sigma_i$, there is a \emph{run} of $\mathcal{A}$ such that $\rho=(q_0,\nu_0)\xrightarrow{t_1,\sigma_1}(q_1,\nu_1)\xrightarrow{t_2,\sigma_2}\cdots\xrightarrow{t_n,\sigma_n}(q_n,\nu_n)$, where $\nu_0(c)=0$, only if (1) $(q_{i-1},\sigma_i,\phi_i,b_i,q_i)\in\Delta$, (2) $\nu_{i-1}(c) + t_{i}$ satisfies $\phi_i$, (3) $\nu_{i}(c)=\nu_{i-1}(c)+t_{i}$ if $b_{i}=\bot$ and $\nu_{i}(c)=0$ otherwise, for all $1\leq i \leq n$. Let $|\omega|$ denote its \emph{length}. When the timed automaton $\mathcal{A}$ is known, each timed word $\omega$ can be extended by including the reset information $b_i$, indicating whether there is a reset after taking the transition for $(\sigma_i,t_i)$. We denote the corresponding \emph{reset timed word} as $\omega_r=(\sigma_1,t_1,b_1)(\sigma_2,t_2,b_2)\cdots(\sigma_n,t_n,b_n)$. If $q_n\in F$, then $\omega$ is an accepting timed word of $\mathcal{A}$. The \emph{(recognized) timed language} $L(\mathcal{A})$  is the set of accepting timed words of $\mathcal{A}$. 

\begin{wrapfigure}{r}{0.25\textwidth}
 \vspace*{-1.2cm}
  \begin{center}
		\resizebox{\linewidth}{!}{
		\begin{tikzpicture}[->,
			>=stealth,
			node distance = 2.5cm,
			every state/.style={thick, fill=gray!10, minimum size=2pt},
			sink/.style={thick, fill=blue!30, minimum size=2pt},
			sink edge/.style={thick, fill=blue!30}
			]
			\node[state, accepting, initial, initial where=above] (q0) {$q_0$};
			\node[state, accepting, right = 2.5cm of q0] (q1) {$q_1$};
			\node[state, sink, below= 1.8cm of $(q0)!0.5!(q1)$] (q2) {$q_2$};
			\draw (q0) edge[bend left, above, black, line width=0.4mm] node{$a, \left[4, 9\right], \bot$} (q1);
			\draw (q1) edge[bend left, below, black, line width=0.4mm] node{$a, \left[4, \infty\right), \bot$} (q0);
			\draw (q0) edge[sloped, anchor=center, below, blue, line width=0.4mm] node{$a, \left[0, 4\right). \top$} (q2);
			\draw (q1) edge[sloped, anchor=center, blue, line width=0.4mm,below] node{$a, \left[0, 4\right), \top$} (q2);
			\draw (q2) edge[loop below, blue, line width=0.4mm] node{$a, \left[0, +\right), \top$} (q2);
			\draw (q0) edge[sloped, below, bend right=50, blue, line width=0.4mm] node{$a, \left(9, \infty\right), \top$} (q2);
		\end{tikzpicture}}
  \end{center}
  \vspace*{-0.5cm}
  \caption{An example of complete DOTA}
  \label{fig:COTA}
\end{wrapfigure}
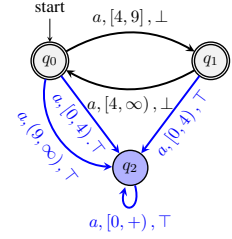

An OTA is a \emph{deterministic one-clock timed automaton} (DOTA) if there is at most one run for any timed word (equivalently, if the guards of transitions from any location under the same action do not intersect each other). Two DOTAs are equivalent if they recognize the same timed language. A DOTA is \emph{complete} if for any $q\in Q$ and action $\sigma\in\Sigma$, the corresponding guards form a partition of $\mathbb{R}_{\geq 0}$. This means any given timed word has exactly one run. Any DOTA $\mathcal{A}$ can be transformed into a complete DOTA (COTA) accepting the same timed language by introducing a non-accepting ``sink'' location and letting all invalid or non-described behaviors go to the sink. We therefore assume that we are working with complete DOTAs. Fig.~\ref{fig:COTA} shows a DOTA $\mathcal{A}$, with $q_2$ added as the sink location to make it complete.

The classic method of active learning is $L^*$ algorithm~\cite{Angluin87} which can be regarded as an interaction between a learner and a teacher, where the teacher can answer membership and equivalence queries. In a membership query, the learner can ask if a word belongs to the target language, or if a constructed automaton accepts that language as an equivalence query. The teacher can answer ``yes'' or ``no'' for the queries and return counterexamples for the equivalence queries. The learner collects query results and counterexample in the observation table consisting of three sets of words: a prefix-closed set $S$, a set $R$ and a suffix-closed set $E$. Each word in $S$ corresponds to a unique location. Words in $R$ can be considered the ``boundary'' of $S$, and $E$ contains suffixes to distinguish different words in $S\cup R$.

An et al.~\cite{AnCZZZ20} described an active learning algorithm for DOTAs. Inherited from $L^{*}$, in a membership query, the teacher is given a timed word. We use $\textsf{MQ}: (\Sigma\times\mathbb{R}_{\geq 0})^*\rightarrow \{+,-\}$ to denote the function mapping timed words to $+$ if it is accepted, and $-$ otherwise. In an equivalence query, the teacher receives a hypothesis DOTA $\mathcal{H}$, and return a counterexample in the negative case in addition. In that work, a ``smart'' teacher is first assumed, who also returns the reset information for any timed word given as membership query. With this reset information, an observation table can be maintained where each row in $S\cup R$ and each column in $E$ is a \emph{logical timed word}, which contains the clock valuation after each transition rather than the delay times. When the observation table satisfies readiness conditions, a candidate automaton is constructed in a manner similar to the learning of symbolic automata~\cite{Drews17}. Next, in the ``normal'' teacher setting, where the teacher does not return reset information, the algorithm searches over all possible choices of resets for timed words in $S\cup R$, constructing one observation table for each. Due to the large time and memory requirements, the algorithm has limited scalability (up to six locations for the randomly constructed examples in the paper).

\section{Learning Algorithm}\label{sec:learning_CS}

In this section, we describe a new algorithm for active learning of DOTAs by incorporating constraint solving using SMT solvers. 
The main idea of the algorithm is to maintain a single observation table that collects all results from previous membership queries, rather than one observation table for each possible choice of resets. Instead, the reset information is encoded as boolean variables with unknown values. At any time, the observation table may be ready for some choice of resets, but not for others. We encode the readiness conditions for the observation table as a formula on the variables for reset information, as well as for the location assignments of rows in the table. An SMT solver is then used to solve these constraints, returning a choice of resets that make the table ready, as well as a candidate automaton that can be sent for equivalence query.

Several difficulties must be addressed in order to realize this idea. In particular, whether two rows in the observation table are distinguishable is no longer clear-cut, but may depend on the choice of resets. This also means the partition of rows into $S$ and $R$ in the traditional $L^*$ algorithm need to be extended, by adding a category $S_+$ of rows that are distinguishable from rows in $S$ only for some choice of resets, that are needed for building candidate automaton.

In the remainder of this section, we first discuss comparison of timed words without being certain of clock-reset information, then extensions made to the observation table, method of encoding readiness constraints, and finally the main algorithm together with termination and complexity analysis.

\subsection{Alignment and comparison of timed words}
\label{sec:alignment-comparison}

In the $L^*$ framework, one key step is to determine if two words $w_1$ and $w_2$ belong to different equivalence classes of the target regular language, i.e., arrive at different locations of the underlying target DFA $A$. It is achieved by membership queries via testing words $w_1$ and $w_2$ after appending some suffix $e$. If $w_1\cdot e$ is an accepting word and $w_2\cdot e$ is not (or vice versa), then $w_1$ and $w_2$ must arrive at different locations.

In our case, suppose $\mathcal{A}$ is the target DOTA to be learned. Given two timed words $\omega_1$ and $\omega_2$, we wish to determine whether they arrive at different locations of $\mathcal{A}$. However, since the reset information is not observable when running timed words $\omega_1$ and $\omega_2$, the values of the clock at the end is unknown. If the values of the clock are not the same, the result of running $\omega_1\cdot e$ and $\omega_2\cdot e$ may be different even if $\omega_1$ and $\omega_2$ arrive at the same location. In order to effectively test using the suffix $e$, we need to suppose that the last time when the clock resets in $\omega_1$ and $\omega_2$ are known, and then align the values of the clock before executing the suffix $e$. We state these concepts more formally in the following definitions.

\begin{definition}[Last reset of a timed word]
Given a timed word $\omega = (\sigma_1,t_1) (\sigma_2,t_2) \cdots \\(\sigma_n,t_n)$, and DOTA $\mathcal{A}$. Let $\omega_r = (\sigma_1,t_1,b_1) (\sigma_2,t_2,b_2) \cdots (\sigma_n,t_n,b_n)$ be the reset timed word that results from running $\omega$ on $\mathcal{A}$. The \emph{last reset} $k_{\mathcal{A}}(\omega)$ is defined to be 0 if $b_i=\bot$ for all $1\leq i\leq n$, and $k_{\mathcal{A}}(\omega) = i$ if $b_i=\top$ and $b_j=\bot$ for all $j > i$.
\end{definition}

Suppose the last reset $k_{\mathcal{A}}(\omega)$ is known for a timed word $\omega$, then we can compute the value of clock after executing $\omega$ on $\mathcal{A}$. Let $\nu_c(\omega,i)$ be the value of clock after executing $\omega$ if the last reset equals $i$. This is computed as the sum of $t_{i+1}$ to $t_n$, where $n$ is the length of $\omega$ (if $i=n$, for the case where the last reset occurs after $(\sigma_n,t_n)$, then $\nu_c(\omega,i)=0$). Hence, given two timed words $\omega_1$ and $\omega_2$ with known values of last resets $i_1,i_2$, it is possible to \emph{align} the two timed words for testing a suffix $e$.

\begin{definition}[Alignment for testing on a suffix]
	\label{def:alignment-testing}
	Consider two timed words $\omega_1,\omega_2$, and suppose their last resets are $i_1,i_2$ respectively. Let $e=(\sigma_1,t_1) (\sigma_2,t_2)\cdots (\sigma_m,t_m)$ be a nonempty suffix. Let $\nu_1=\nu_c(\omega_1,i_1)$ and $\nu_2=\nu_c(\omega_2,i_2)$. Then form the suffixes $e_1,e_2$ depending on the following cases:
	\begin{itemize}
		\item If $\nu_1 > \nu_2$, then let $e_1 = e$ and $e_2 = (\sigma_1, t_1+(\nu_1-\nu_2))\cdot(\sigma_2,t_2)\cdots (\sigma_m, t_m)$.
		\item If $\nu_1 < \nu_2$, then let $e_1 = (\sigma_1, t_1+(\nu_2-\nu_1))\cdot(\sigma_2,t_2)\cdots (\sigma_m,t_m)$ and $e_2 = e$.
		\item If $\nu_1 = \nu_2$, then let $e_1 = e_2 = e$.
	\end{itemize}
	Define a \emph{test} $T(\omega_1,\omega_2,i_1,i_2,e)$ between $\omega_1$ and $\omega_2$ with suffix $e$ and last resets $i_1,i_2$ as follows. If $e$ is nonempty, then the test compares results for two membership queries $\omega_1\cdot e_1$ and $\omega_2\cdot e_2$. The test succeeds, i.e. $T(\omega_1,\omega_2,i_1,i_2,e) = \top$, if $\textsf{MQ}(\omega_1\cdot e_1) = \textsf{MQ}(\omega_2\cdot e_2)$. Otherwise $T(\omega_1,\omega_2,i_1,i_2,e) = \bot$. If $e$ is empty, the test simply compares the results for membership queries $\omega_1$ and $\omega_2$.
\end{definition}

It is clear that with the definition of $e_1$ and $e_2$, the value of clock when executing the first timed action of $e_1$ and $e_2$ during the tests must be the same. Hence, if $\omega_1$ and $\omega_2$ arrive at the same location, then the behavior on $e_1$ and $e_2$ must be the same as well. Hence, we obtain the following lemma.

\begin{lemma}[Distinguishable timed words]
\label{lemma:dtw}
If the test $T$ between $\omega_1$ and $\omega_2$ with suffix $e$ and last resets $i_1, i_2$ fails, then for any DOTA $\mathcal{A}$ such that $i_1=k_{\mathcal{A}}(\omega_1)$ and $i_2=k_{\mathcal{A}}(\omega_2)$, the timed words $\omega_1$ and $\omega_2$ must arrive at different locations in $\mathcal{A}$.
\end{lemma}

\begin{example}\label{expl:align}
Consider two timed words $\omega_1 = (a,4)$ and $\omega_2 = \epsilon$. They are both accepted by $\mathcal{A}$ in Fig.~\ref{fig:COTA}. Consider suffix $e=(a,5.5)$. If the last resets are $i_1=0$ and $i_2=0$ (the correct reset for $\mathcal{A}$), then $e_1=(a,5.5)$ and $e_2=(a,9.5)$, and $\textsf{MQ}(\omega_1\cdot e_1)=+$, $\textsf{MQ}(\omega_2\cdot e_2)=-$, so they can be distinguished, and indeed $\omega_1$ and $\omega_2$ arrive at different locations as required by Lemma~\ref{lemma:dtw}. If the last resets are $i_1=1$ and $i_2=0$, then $e_1=e_2=(a,5.5)$, and $\textsf{MQ}(\omega_1\cdot e_1)=\textsf{MQ}(\omega_2\cdot e_2)=+$, so they cannot be distinguished. Hence, when we do not know the true reset information, whether $\omega_1$ and $\omega_2$ are distinguishable by $e=(a,5.5)$ depends on the choice of resets. \demo
\end{example}

Since $\mathcal{A}$ is unknown during learning, we need to iterate over all possible combinations of last resets. For two timed words $\omega_1$ and $\omega_2$, we define the set of valid combinations of last resets as follows.

\begin{definition}[Valid combinations of last resets]
Consider two timed words $\omega_1,\omega_2$ and suppose the length of the longest common prefix of $\omega_1$ and $\omega_2$ is $m$. Then the set $\mathcal{C}(\omega_1,\omega_2)$ of valid combinations of last resets is
\[ \mathcal{C}(\omega_1,\omega_2) = \{(i_1,i_2)\,\vert\,0\le i_1\le |\omega_1| \wedge 0\le i_2\le |\omega_2| \wedge (i_1\le m \wedge i_2\le m \Rightarrow i_1 = i_2) \} \]
\end{definition}

\begin{example}
Suppose $\omega_1=(a,4)$ and $\omega_2=(a,4)(a,5.5)$, then $m=1$, and the set of valid combinations are 
$\mathcal{C}(\omega_1,\omega_2) = \left\{(0,0),(0,2),(1,1), (1,2)\right\}$. In particular $(0,1)$ or $(1,0)$ are not allowed, since they give contradicting reset choices for the transition taken by $(a,4)$. \demo
\end{example}

\subsection{Timed observation table}\label{sbsc:table}

\begin{definition}[Observation table]\label{def:table}
An observation table $\mathcal{O}=(S, S_+, R, E, f, N)$ is a 6-tuple, satisfying the following conditions:
\begin{itemize}
\item $S, S_+, R$ are disjoint finite sets of timed words called \emph{prefixes}. $S\cup S_+\cup R$ is prefix-closed and $\epsilon\in S$. If $\omega\in S\cup S_+$ and $\sigma\in\Sigma$, then $\omega\cdot (\sigma,0)\in S\cup S_+\cup R$.
	
\item $E$ is a finite set of timed words called \emph{suffixes}, with $\epsilon\in E$.
	
\item $f$ is a function mapping pairs $\omega_1,\omega_2\in S\cup S_+\cup R$ and $(i_1,i_2)\in \mathcal{C}(\omega_1,\omega_2)$ to $\mathbb{B}$, indicating whether $\omega_1$ and $\omega_2$ are currently distinguished under last resets $i_1,i_2$.
	
\item $N$ is the current limit on the number of locations in the candidate automaton.
\end{itemize}
\end{definition}

The value $f(\omega_1,\omega_2,i_1,i_2)$ is computed as follows. For each suffix $e\in E$, test the pair $\omega_1,\omega_2$ under last resets $i_1,i_2$ and suffix $e$ as in Definition~\ref{def:alignment-testing}. If $T(\omega_1,\omega_2,i_1,i_2,e)=\top$ for all $e\in E$, then $f(\omega_1,\omega_2,i_1,i_2)=\top$, otherwise $f(\omega_1,\omega_2,i_1,i_2)=\bot$. 

\begin{definition}[Certainly distinct rows]\label{def:distinct}
Given an observation table $\mathcal{O}$, two rows $\omega_1,\omega_2$ are \emph{certainly distinct} if $f(\omega_1,\omega_2,i_1,i_2)=\bot$ for all $i_1,i_2\in\mathcal{C}(\omega_1,\omega_2)$.
\end{definition}

We first explain the different components of $\mathcal{O}$ in an intuitive way. The set $S$ contains timed words that are certainly distinct from each other. The set $S_+$ are additional rows in the observation table that are distinct from rows in $S$ under some choice of resets, that are required to be in the interior for some candidate automata. The set $R$ is the boundary of the observation table as usual for $L^*$ algorithms. The condition that $\omega\cdot(\sigma,0)\in S\cup S_+\cup R$ is analogous to the condition that $\omega\cdot \sigma\in S\cup R$ in the DFA case. It enforces that information is available to construct the transitions in the candidate automaton for each location and action $\sigma\in\Sigma$.

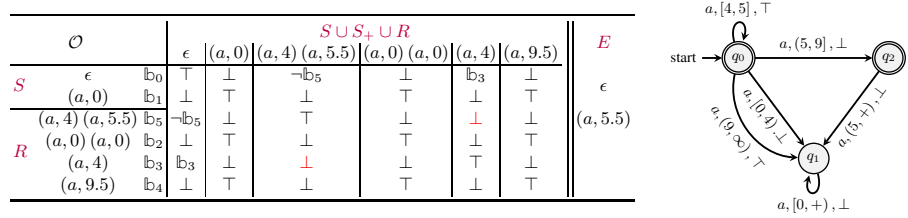
\begin{figure}
\vspace*{-0.5cm}
\centering
\begin{minipage}{0.69\linewidth}
\centering
    \resizebox{\linewidth}{!}{
    \begin{tabular}{lccc|c|c|c|c|c|c c||c}
		   \toprule
		   \multicolumn{3}{c}{\multirow{2}{*}{$\mathcal{O}$}} & &\multicolumn{6}{c}{\textcolor{purple}{$S\cup S_+ \cup R$}} & & \multirow{2}*{\textcolor{purple}{$E$}}\\
		   & & & & $\epsilon$ &$\left(a, 0\right)$ & $\left(a,4\right)\left(a,5.5\right)$ & $\left(a,0\right)\left(a, 0\right)$ & $\left(a,4\right)$ & $\left(a, 9.5\right)$ & & \\
		   \cline{1-10}
		   \multirow{2}*{\textcolor{purple}{$S$}} & & $\epsilon$ & \multicolumn{1}{c|}{$\mathbb{b}_0$} & \multicolumn{1}{c|}{$\top$} & \multicolumn{1}{c|}{$\bot$} & \multicolumn{1}{c|}{$\neg \mathbb{b}_5$} & \multicolumn{1}{c|}{$\bot$} & \multicolumn{1}{c|}{$\mathbb{b}_3$} & $\bot$ & & \multirow{2}*{$\epsilon$}\\
		   & & $(a, 0)$ & \multicolumn{1}{c|}{$\mathbb{b}_1$} &\multicolumn{1}{c|}{$\bot$} & \multicolumn{1}{c|}{$\top$} & \multicolumn{1}{c|}{$\bot$} & \multicolumn{1}{c|}{$\top$} & \multicolumn{1}{c|}{$\bot$} & $\top$ & & \\
		   \cline{1-4}
		   \multirow{4}*{\textcolor{purple}{$R$}}
		   & & $\left(a, 4\right)\left(a,5.5\right)$ & $\mathbb{b}_5$  & \multicolumn{1}{c|}{$\neg \mathbb{b}_5$} & \multicolumn{1}{c|}{$\bot$} & \multicolumn{1}{c|}{$\top$} & \multicolumn{1}{c|}{$\bot$} & \multicolumn{1}{c|}{\textcolor{red}{$\bot$}} & $\bot$ & & $(a,5.5)$\\
		   & & $\left(a, 0\right)\left(a, 0\right)$ & $\mathbb{b}_2$ & \multicolumn{1}{c|}{$\bot$} & \multicolumn{1}{c|}{$\top$} & \multicolumn{1}{c|}{$\bot$} & \multicolumn{1}{c|}{$\top$} & \multicolumn{1}{c|}{$\bot$} & $\top$ & & \\
		   & & $\left(a, 4\right)$ & $\mathbb{b}_3$  & \multicolumn{1}{c|}{$\mathbb{b}_3$} & \multicolumn{1}{c|}{$\bot$} & \multicolumn{1}{c|}{\textcolor{red}{$\mathbf{\bot}$}} & \multicolumn{1}{c|}{$\bot$} & \multicolumn{1}{c|}{$\top$} & $\bot$ & & \\
		   & & $\left(a, 9.5\right)$ & $\mathbb{b}_4$  & \multicolumn{1}{c|}{$\bot$} & \multicolumn{1}{c|}{$\top$} & \multicolumn{1}{c|}{$\bot$} & \multicolumn{1}{c|}{$\top$} & \multicolumn{1}{c|}{$\bot$} & $\top$ & & \\
		  \bottomrule
		\end{tabular}
	}
\end{minipage}
\hfill
\begin{minipage}{0.3\linewidth}
\centering
    \resizebox{0.9\linewidth}{!}{
        \begin{tikzpicture}[->,
			>=stealth,
			node distance = 2.5cm,
			line width=0.4mm,
			every state/.style={thick, fill=gray!10, minimum size=2pt},
			sink/.style={thick, fill=blue!30, minimum size=2pt},
			sink edge/.style={thick, fill=blue!30}
			]
			\node[state, accepting, initial, initial where=left] (q0) {$q_0$};
			\node[state, accepting, right = 2.5cm of q0] (q2) {$q_2$};
			\node[state, below= 1.8cm of $(q0)!0.5!(q2)$] (q1) {$q_1$};
			\draw (q0) edge[loop above] node{$a, \left[4, 5\right], \top$} (q0);
			\draw (q0) edge[above, black] node{$a, \left(5, 9\right], \bot$} (q2);
			\draw (q0) edge[sloped, anchor=center, below, line width=0.4mm] node{$a, \left[0, 4\right). \bot$} (q1);
			\draw (q2) edge[sloped, anchor=center, below] node{$a, \left(5, +\right), \bot$} (q1);
			\draw (q1) edge[loop below, line width=0.4mm] node{$a, \left[0, +\right), \bot$} (q1);
			\draw (q0) edge[sloped, below, bend right=50] node{$a, \left(9, \infty\right), \top$} (q1);
		\end{tikzpicture}
		}
\end{minipage}
\vspace*{-0.3cm}
\caption{\textbf{Left}: an instance of observation table $\mathcal{O}$ during learning of $\mathcal{A}$ in Fig.~\ref{fig:COTA}; \textbf{Right}: candidate DOTA constructed from the table after moving $(a,4)(a,5.5)$ to $S_+$.}
\label{fig:table_and_dota}
\vspace*{-0.9cm}
\end{figure}

\begin{example}\label{ex:s+}
\textbf{}Fig.~\ref{fig:table_and_dota} shows an instance of the observation table. The rows of the table are timed words in $S,S_+$ and $R$ (here $S_+$ is empty). The columns of the table are indexed by the same timed words. Each cell in the table summarizes when two timed words are distinguished (the function $f$), using formulas in terms of reset variables $\mathbb{b}$ introduced in Section~\ref{sec:encoding}. The table also shows the list of suffixes $E$. For example, for timed words $(a,4)$ and $\epsilon$, we need to compare them on suffix $(a,5.5)$ after alignment. According to the analysis in Example~\ref{expl:align}, we have $f((a,4),\epsilon,0,0)=\bot$ and $f((a,4),\epsilon,1,0)=\top$. This is summarized as the expression $\mathbb{b}_3$. \demo

\end{example}

The main difference between the observation table defined here and the usual ones for $L^*$ algorithms is that we do not record a particular query result for each prefix in $S\cup S_+\cup R$ and suffix $E$, as these results cannot be used effectively without knowing reset information. In contrast, we maintain which pairs of rows can be distinguished for each possible choice of reset information. This can be contrasted with the approach in~\cite{AnCZZZ20}. Rather than maintaining a copy of the observation table for each combination of reset information, our method records all information obtained so far in a single table, with reset information determined by constraint solving.

\subsection{Encoding of readiness constraints}
\label{sec:encoding}

To obtain a hypothesis DOTA $\mathcal{H}$ from observation table $\mathcal{O}$, we should provide an assignment for the location and reset information of each timed word in $\ssr$, which ensure readiness conditions for the table, such as closedness and consistency. The main idea is to encode such readiness conditions as formulas in terms of location and reset assignments, and then use SMT solvers to find feasible solutions to these constraints or prove that they are not satisfiable. The constraints are stated in terms of two variables for each row $\omega\in\ssr$: an \emph{ending reset variable} $\bv_\omega$ and a \emph{location variable} $\qv_\omega$.

\begin{definition}[Ending reset variable and location variable]
Given a timed word $\omega = (\sigma_1, t_1)\dots (\sigma_n, t_n)\in\ssr$, define the \emph{ending reset variable} $\bv_\omega\in\{\top,\bot\}$ to denote whether clock resets after running the final timed action $(\sigma_n,t_n)$ (for the empty timed word $\epsilon$, we declare $\bv_\epsilon=\top$). Define the \emph{location variable} $\qv_\omega\in\{1,\dots,N\}$ to represent the location after running $\omega$ in the candidate automaton.
\end{definition}

Since the set of timed words $\ssr$ is prefix-closed, the ending reset variables $\{\bv_\omega\}_{\omega\in\ssr}$ in fact determine whether the clock resets after each timed action for each row in $\ssr$. In particular, we can encode the last reset for $\omega$ in terms of the ending reset variables.

\begin{definition}[Encoding of last reset]
Given $\omega = (\sigma_1, t_1)...(\sigma_n, t_n)\in\ssr$. Let $\omega\vert_i$ for $0\le i\le n$ be the prefix of $\omega$ with length $i$. Since $\ssr$ is prefix-closed, we have each $\omega\vert_i\in\ssr$ as well. Let $lr(\omega,i)$, encoding the condition that the last reset of $\omega$ equals $i$, be defined as follows.
\[ lr(\omega,i) \ \triangleq \ \bv_{\omega\vert_i} \wedge \bigwedge_{i<j\le n} \neg \bv_{\omega\vert_j}. \]
For each pair of rows $\omega_1,\omega_2\in\ssr$ and each pair of last resets $(i,j)\in\mathcal{C}(\omega_1,\omega_2)$, the condition that the last resets of $\omega_1,\omega_2$ equal $i,j$ respectively is encoded as follows.
\[LR(\omega_1,\omega_2,i,j) \ \triangleq \ lr(\omega_1, i)\wedge lr(\omega_2, j)\]
\end{definition}

Based on the encoding of last reset, the readiness constraints for the observation table can be encoded in terms of the above variables as follows.

\begin{constraint}[Distinctness of rows]
\label{con1}
\normalfont
Given timed words $\omega_1,\omega_2\in \ssr$, and last resets $i,j\in\mathcal{C}(\omega_1,\omega_2)$, suppose $f(\omega_1, \omega_2, i, j)=\bot$ (meaning $\omega_1$ and $\omega_2$ can be distinguished under last resets $i,j$ in the observation table), then we have the following constraint, indicating $\omega_1$ and $\omega_2$ cannot be assigned to the same location.
\[ C_1(\omega_1,\omega_2,i,j) \ \triangleq\  LR(\omega_1, \omega_2, i, j) \Rightarrow \qv_{\omega_1} \neq \qv_{\omega_2} \]
\end{constraint}
Define the constraint $C_1$ to be the conjunction of all $C_1(\omega_1,\omega_2,i,j)$, for all pairs of rows and valid last resets that can be distinguished.
\[ C_1 \ \triangleq\  \bigwedge_{\substack{\omega_1,\omega_2\in\ssr,\\(i,j)\in\mathcal{C}(\omega_1,\omega_2),\\f(\omega_1,\omega_2,i,j)=\bot}} C_1(\omega_1,\omega_2,i,j). \]

\begin{constraint}[Consistency]\label{con2}
\normalfont
Given timed words $\omega_1, \omega_2\in \ssr$ and last resets $i,j\in\mathcal{C}(\omega_1,\omega_2)$. Suppose $\omega_1'=\omega_1\cdot(\sigma, t_1)$ and $\omega_2'=\omega_2\cdot (\sigma, t_2)$ also appear in $\ssr$, for some $\sigma\in\Sigma$ and $t_1,t_2\in\mathbb{R}_{\ge 0}$. Suppose that under the last resets $i,j$, the value of clock after executing last timed action of $\omega_1'$, but before possible resets, is in the same region as that for $\omega_2'$, then if $\omega_1$ and $\omega_2$ also go to the same location, the transition to be carried out for the last timed action of $\omega_1'$ and $\omega_2'$ must be the same. Hence both ending reset and location for $\omega_1'$ and $\omega_2'$ must be the same. This is encoded as constraints as follows.
\[ C_2(\omega_1,\omega_2,i,j,\sigma,t_1,t_2) \ \triangleq \ 
\qv_{\omega_1} = \qv_{\omega_2} \wedge LR(\omega_1,\omega_2,i,j) \Rightarrow \bv_{\omega_1'} = \bv_{\omega_2'} \wedge \qv_{\omega_1'} = \qv_{\omega_2'}.\]
It is added as a constraint only if $\llbracket\nu_c(\omega_1,i)+t_1\rrbracket = \llbracket\nu_c(\omega_2,j)+t_2\rrbracket$, and if $f(\omega_1,\omega_2,i,j) = \top$. We define constraint $C_2$ to be the conjunction of all such constraints.
\[ C_2\ \triangleq\ \bigwedge_{\substack{\omega_1,\omega_2\in\ssr,\\
\omega_1\cdot(\sigma,t_1), \omega_2\cdot(\sigma,t_2)\in\ssr,\\
(i,j)\in\mathcal{C}(\omega_1,\omega_2),
f(\omega_1,\omega_2,i,j)=\top,\\ \llbracket\nu_c(\omega_1,i)+t_1\rrbracket= \llbracket\nu_c(\omega_2,j)+t_2\rrbracket}} C_2(\omega_1,\omega_2,i,j,\sigma,t_1,t_2). \]
\end{constraint}

\begin{constraint}[Closedness]\label{con3}
\normalfont
The closedness condition for usual $L^*$ algorithms states that each row in $R$ must be represented by a row in $S$. In our case, we require that each row in $R$ is represented by a row in $S\cup S_+$. This translates to the constraint that each location in the candidate automaton must be represented by a row in $S\cup S_+$, encoded as follows (recall $N$ is the current limit on the number of locations).
\[ C_3 \ \triangleq\ \bigwedge_{1\le i\le N} \bigvee_{\omega\in S\cup S_+} \qv_\omega = i \ \wedge \  C_3' \mbox{ where } C_3' \ \triangleq\ \bigwedge_{\omega\in\ssr} 1\le \qv_\omega \le N. \]
During the algorithm, we also make constraint solving queries where the closedness condition is not enforced. Then only the second part $C_3'$ is used.
\end{constraint}

\begin{constraint}[Special assignments]\label{con4}
\normalfont
In order to speed-up constraint solving, we directly make assignments to the location variables of rows in $S$. Order the rows of $S$ as $S=\{\omega_1,\omega_2,\dots,\omega_{|S|}\}$, then the special assignments are encoded as follows.
\[ C_4 \ \triangleq\ \bigwedge_{1\le i\le |S|} \qv_{\omega_i} = i. \]
\end{constraint}

In the main learning algorithm, we will use SMT solvers to attempt to find solutions to these constraints. The algorithm will first attempt to find a solution using $C_3$ together with $C_1,C_2$ and $C_4$. If a solution is found, it proceeds to hypothesis construction as described in Section~\ref{sec:hypothesis-construction}. Otherwise, it attempts to find a solution using $C_3'$ or by increasing $N$. The details are described in Section~\ref{sec:main-algorithm}.

\subsection{Hypothesis construction}
\label{sec:hypothesis-construction}

Once the SMT solver gives a model satisfying constraints in Section~\ref{sec:encoding}, we can build a hypothesis DOTA $\mathcal{H}=(\Sigma,Q_{\mathcal{H}},q^{\mathcal{H}}_0,F_{\mathcal{H}},c,\Delta_{\mathcal{H}})$ from observation table $\mathcal{O}=(S, S_+, R,\\ E, f, N)$ and assignments $\overline{\bv_\omega}$ and $\overline{\qv_\omega\vphantom{\bv_\omega}}$ to ending reset variable and location variables in the model. We define location set $Q_{\mathcal{H}}=\{\overline{\qv_\omega\vphantom{\bv_\omega}}\mid\omega \in S\cup S_+\}$, initial location $q^{\mathcal{H}}_0=\overline{\qv_{\epsilon}\vphantom{\bv_\omega}}$, and accepting locations $F_{\mathcal{H}}=\{\overline{\qv_{\omega}\vphantom{\bv_\omega}}\mid\textsf{MQ}(\omega)=+ \wedge \omega\in S\cup S_+\}$. Next, we describe how to construct the transitions $\Delta_{\mathcal{H}}$.

Given two rows $\omega_1,\omega_2\in\ssr$ such that $\omega_2 = \omega_1 \cdot (\sigma, t)$, we can construct an auxiliary transition $\delta' = (\overline{\qv_{\omega_1}\vphantom{\bv_{\omega_2}}}, \sigma, \psi ,\overline{\bv_{\omega_2}} ,\overline{\qv_{\omega_2}\vphantom{\bv_\omega}})$ with $\psi=\nu(\omega_1) + t$ where $\nu(\omega_1)$ is the value of clock after executing $\omega_1$. Since the table is prefix-closed, the reset information for every timed action in $\omega_1$ has been determined. Therefore, we can determine $\nu(\omega_1)$. We collect all such auxiliary transitions as the set $\Delta'$. 

For any $q \in Q_{\mathcal{H}}$ and $\sigma \in \Sigma$, let $\Psi_{q,\sigma} = \{\psi \mid (q, \sigma, \psi, b, q') \in \Delta' \}$ be the list of clock values on auxiliary transitions from $q$ and with action $\sigma$. We sort $\Psi_{q,\sigma}$ and apply the partition function $P(\cdot)$ to obtain $m$ intervals, written as $g_1,\cdots, g_m$, satisfying $\psi_i \in g_i$ for any $1\leq i \leq m$, where $m=\lvert\Psi_{q,\sigma}\rvert$; 
consequently, for every $(q, \sigma, \psi_i, b, q') \in \Delta'$, a transition $\delta_i = (q,\sigma,g_i, b, q')$ is added to $\Delta_{\mathcal{H}}$. This determines the transitions between locations in $\mathcal{H}$ and hence finishes the construction. The partition function $P(\cdot)$ is taken from~\cite{AnCZZZ20}, and also similar to that used for learning symbolic automata~\cite{Drews17}. Note the condition $\omega\cdot(\sigma,0)\in S\cup S_+\cup R$ in Definition~\ref{def:table} enforces $\mu_0=0$ below.

\begin{definition}[Partition function]\label{def:partition}
Given a list of clock valuations $\ell =\mu_0,\mu_1,\cdots,\mu_n$ with 
$0=\mu_0 < \mu_1 \cdots < \mu_n$, 
and $\lfloor \mu_i \rfloor \neq \lfloor \mu_j \rfloor$ if $\mu_i,\mu_j\in\mathbb{R}_{\geq 0}\setminus\mathbb{N}$ and $i\neq j$ 
for all $1\leq i, j\leq n$, let $\mu_{n+1} = 	 \infty$, then a partition function $P(\cdot)$ mapping $\ell$ to a set of intervals $\{g_0,g_1,\dots,g_n\}$, 
which is a partition of $\mathbb{R}_{\geq 0}$, is defined as  
\begin{equation*}
\footnotesize
	g_i = \begin{cases}
		[\mu_i, \mu_{i+1}), & \text{if}\ \  \mu_i \in \mathbb{N} \wedge \mu_{i+1} \in \mathbb{N}; \\
		(\lfloor \mu_i \rfloor, \mu_{i+1}), & \text{if}\ \  \mu_i \in \mathbb{R}_{\geq 0}\setminus\mathbb{N} \wedge \mu_{i+1} \in \mathbb{N}; \\
		[\mu_i, \lfloor\mu_{i+1}\rfloor], & \text{if}\ \  \mu_i \in \mathbb{N} \wedge \mu_{i+1} \in \mathbb{R}_{\geq 0}\setminus\mathbb{N}; \\
		(\lfloor\mu_i\rfloor,\lfloor\mu_{i+1}\rfloor], & \text{if}\ \  \mu_i \in \mathbb{R}_{\geq 0}\setminus\mathbb{N} \wedge \mu_{i+1} \in \mathbb{R}_{\geq 0}\setminus\mathbb{N}.
	\end{cases}
\end{equation*}
\end{definition}

Since the table $\mathcal{O}$ with the feasible assignments satisfies the readiness constraints, the constructed hypothesis is a deterministic one-clock timed automaton, and agrees with accepting information for rows in $\ssr$. This is stated as the following theorem.

\begin{theorem}
\label{thm:hypothesis}
Given observation table $\mathcal{O}=(S, S_+, R, E, f, N)$ and feasible assignments to $\overline{\bv_\omega}$ and $\overline{\qv_\omega\vphantom{\bv_\omega}}$, the hypothesis $\mathcal{H} = (\Sigma,Q,q_0,F,c,\Delta)$ is deterministic. For each row $\omega\in\ssr$,
$\mathcal{H}$ accepts the timed word $\omega$ iff $\textsf{MQ}(\omega) = +$. Finally, for any two rows $\omega_1,\omega_2\in \ssr$, if the value of $f$ on $\omega_1,\omega_2$ and the setting of reset variables $\overline{\bv}$ is $\bot$, then $\overline{\qv_{\omega_1}\vphantom{\bv_{\omega_1}}}\neq \overline{\qv_{\omega_2}\vphantom{\bv_{\omega_2}}}$, and the two rows reach distinct locations in $\mathcal{H}$.
\end{theorem}

After the hypothesis $\mathcal{H}$ is built, it is sent for an equivalence query. If the teacher returns a counterexample $ctx$, the learner adds all prefixes of $ctx$ to $R$ during counterexample processing.

\subsection{Main algorithm and correctness}
\label{sec:main-algorithm}

\begin{algorithm}[!t]
\caption{Learning DOTA using constraint solving}
\label{alg:learning}
\SetKwInOut{Input}{input}
\SetKwInOut{Output}{output}
\SetKw{Continue}{continue}
\Input{an observation table $\mathcal{O}=(S, S_+, R, E, f, N)$,
	the alphabet $\Sigma$.}
\Output{an automata $\mathcal{H}$ recognizing the target language $L$.}
$S\leftarrow \{\epsilon\}$, $S_{+}\leftarrow \emptyset$, $R\leftarrow\{(\sigma,0) \mid \sigma \in \Sigma \}$, $E\leftarrow \{\epsilon\}$, $N\leftarrow 1$ \tcp*{initialization}
\While{$\top$}{
	$\mathcal{O}\leftarrow \mathsf{move\_to\_S}(\mathcal{O})$\; \label{line:move_to_s}
	$\mathit{flag}$, $M$ $\leftarrow$ $\mathsf{SMT}$($C_1\wedge C_2\wedge C_3 \wedge C_4$) \tcp*{solve constraints to get model $M$} \label{line:SMT_cons}
	\If{$\mathit{flag}=\top$}{ 
		$\mathcal{H}\leftarrow$ $\mathsf{build\_hypothesis}$($\mathcal{O}$, $M$) \tcp*{build $\mathcal{H}$  from table $\mathcal{O}$ and model $M$} \label{line:hypo}
		$\mathit{equivalent}$, $\mathit{ctx}$ $\leftarrow$ $\mathsf{equivalence\_query}$($\mathcal{H}$)\;
		\If{$\mathit{equivalent}=\top$}{
			\Return $\mathcal{H}$ \tcp*{success} \label{line:return}
		}
		\Else{
			$\mathcal{O}\leftarrow $ $\mathsf{ctx\_processing}$($\mathcal{O}$, $\mathit{ctx}$) \tcp*{counterexample processing} \label{line:ctx}
		}
	}
	\Else{ 
		$\mathit{flag}$, $M'$ $\leftarrow$ $\mathsf{SMT}$($C_1\wedge C_2\wedge C_3' \wedge C_4$) \tcp*{solve relaxed constraints} \label{line:relax_cons}
		\If{$\mathit{flag}=\top$}{
			$\mathcal{O}$ $\leftarrow $ $\mathsf{move\_to\_S_{+}}$($\mathcal{O}$, $M'$) \tcp*{modify table $\mathcal{O}$ guided by solution $M'$} \label{line:modify}
		}
		\Else{
			$N\leftarrow N+1$ \tcp*{try for larger number of locations} \label{line:increase}
		}
	}
}
\end{algorithm}

The overall procedure of the algorithm is given in Algorithm~\ref{alg:learning}. The observation table $\mathcal{O}=(S, S_+, R, E, f, N)$ is initialized with $S=\{\epsilon\}$, $S_{+}=\emptyset$, $R=\{(\sigma, 0)\mid \sigma \in \Sigma\}$, $E=\{\epsilon\}$, and $N=1$. The function $\mathsf{move\_to\_S}$ tests each row in $R$ to see if it is certainly distinct (according to Definition~\ref{def:distinct}) from each row in $S$. If so the certainly distinct row is moved to $S$. For each row $\omega$ moved to $S$, $\omega\cdot (\sigma,0)$ is added to $R$ for every $\sigma\in\Sigma$ (Line~\ref{line:move_to_s}). After that, the formula $C_1\wedge C_2\wedge C_3\wedge C_4$ is built and sent to an SMT solver (Line~\ref{line:SMT_cons}). If a solution $M$ is found for the ending reset and location variables, then a hypothesis $\mathcal{H}$ is constructed from the table $\mathcal{O}$ and the solution $M$ (Line~\ref{line:hypo}), and an equivalence query is performed to determine whether the hypothesis $\mathcal{H}$ is correct. If the answer is positive, the algorithm returns with automaton $\mathcal{H}$ (Line~\ref{line:return}). Otherwise, the learner updates the table $\mathcal{O}$ by adding all prefixes of the returned counterexample $\mathit{ctx}$ to $R$ (Line~\ref{line:ctx}), and begins a new iteration starting from finding certainly distinct rows (Line~\ref{line:move_to_s}) and updating constraints.
Note that new suffixes may be added to $E$ during the computation of constraint $C_2$. If two timed words $\omega_1\cdot (\sigma, t_1)$ and $\omega_2\cdot (\sigma, t_2)$ end in the same region under some choice of resets, $\omega_1$ and $\omega_2$ are currently indistinguishable under this choice, but $\omega_1\cdot (\sigma,t_1)$ and $\omega_2\cdot (\sigma,t_2)$ can be distinguished with suffix $e\in E$, then the timed word $(\sigma,\mathsf{min}(t_1, t_2))\cdot e$ is added to $E$. This allows us to distinguish $\omega_1$ and $\omega_2$ directly using $(\sigma,\mathsf{min}(t_1,t_2))\cdot e\in E$. After new suffixes are added to $E$, the entire observation table need to be updated, with possible new distinguishable pairs and new rows added to $S$.

If there is no solution to $C_1\wedge C_2\wedge C_3\wedge C_4$, then the learner first relaxes $C_3$ to $C_3'$ (Line~\ref{line:relax_cons}). It is now permitted that some rows in $R$ are assigned to a location different from any row in $S\cup S_{+}$. If there is a solution for the relaxed condition, it indicates that some row in $R$ may represent a new location, even though it is not certainly distinct from all rows in $S$. The function $\mathsf{move\_to\_S_+}$ moves such rows from $R$ to $S_{+}$, and add $\omega\cdot (\sigma,0)$ to $R$ for each $\sigma\in\Sigma$ and each $\omega$ moved to $S_{+}$ (Line~\ref{line:modify}). If there is no solution even for the relaxed constraints, the learner increases $N$ by 1 (Line~\ref{line:increase}), attempting to find a model with larger size.

\begin{example}
\label{expl:algorithm}
In the observation table in Fig.~\ref{fig:table_and_dota}, each cell at row $\omega_1$ and column $\omega_2$ records at which choice of resets $\omega_1$ and $\omega_2$ \emph{cannot} be distinguished, as an expression in terms of $\mathbb{b}_\omega$'s. For example, the expression $\top$ means $\omega_1$ and $\omega_2$ cannot be distinguished for all choice of resets, while $\bot$ means $\omega_1$ and $\omega_2$ are certainly distinct.

Constraint~\ref{con3} (Closedness) requires that each row in $R$ is represented by some row in $S\cup S_+$. Although $\omega_3=(a,4)$ or $\omega_5=(a,4)(a,5.5)$ can be represented by $\epsilon$ by setting $\mathbb{b}_3=\top$ or $\mathbb{b}_5=\bot$, they are known to be certainly distinct from each other (indicated by the red $\bot$ in the table), so they cannot be both represented by $\epsilon$. This means $C_1\wedge C_2\wedge C_3\wedge C_4$ is not satisfiable, so we relax the constraint to $C_1\wedge C_2\wedge C_3'\wedge C_4$. This is solvable by setting $\mathbb{b}_3=\top$, and let $\omega_5$ represent an additional location. Then $\omega_5$ is moved to $S_+$, and the candidate DOTA at the right of Fig.~\ref{fig:table_and_dota} can be constructed from the updated observation table. \demo
 
\end{example}

\paragraph{Analysis of the algorithm}

The algorithm is sound since it returns an automaton only if it passes equivalence queries. The termination of the algorithm can be explained through a comparison with a brute-force search version of the algorithm similar to the normal teacher case in~\cite{AnCZZZ20}. The brute-force search constructs a binary tree of observation tables, with each branching corresponds to a choice of reset for some row in $S\cup R$ (here $S_+$ is not needed since reset information is now certain). Our algorithm simulates a breadth-first search on the tree based on the number of locations. Rows in $S_+$ can be viewed as rows that are added to $S$ in some (but not all) branches of the search tree. A simple estimate for the number of rows in $S\cup S_{+}$, and hence in $R$ gives an exponential worst-case bound in terms of $N$. However, in practice it usually increases slowly as shown in our experiments.

\begin{theorem}[Correctness and termination]
\label{thm:correctness}
Algorithm~\ref{alg:learning} always terminates and returns a correct DOTA recognizing the underlying target timed language.
\end{theorem}

\section{Extension to Deterministic Timed Mealy Machines}
\label{sec:mealy}
For practical applications on real-time reactive systems with input/output behavior, we consider a timed version of Mealy machines. Inspired by Mealy machine with one timer (MM1T)~\cite{VaandragerB021}, we divide the actions in $\Sigma$ into \emph{input} and \emph{output} actions. The special \emph{empty action} $\epsilon$ represents the invisible action or nothing happening. We assume that there is a pair of input and output actions on each transition. Hence, the model can also be viewed as a Mealy machine with one clock. 
\begin{definition}[Timed Mealy Machines] \label{def:ocmm}
A timed Mealy machine (TMM) is a 6-tuple $\mathcal{M}=(Q, I, O, q_0, c, \Delta)$, where $Q$ is a finite set of locations; $I$ is a finite set of inputs, containing the special empty action $\epsilon$; $O$ is a finite set of outputs, containing the special empty action $\epsilon$; $q_0$ is the unique initial location; $c$ is the single clock; and $\Delta \subseteq Q\times I \times O \times \Phi_c \times \mathbb{B} \times Q$ is a finite set of transitions.
\end{definition}
A transition $\delta = (q,i,o,\phi,b,q')$ allows a jump from $q$ to $q'$ and generates an output $o$ when provided input $i\in I$ and if $\phi \in \Phi_c$ is satisfied. Meanwhile, clock $c$ is reset to zero if $b = \top$, and remains unchanged otherwise. Given a timed word over inputs $\omega=(i_1,t_1)(i_2,t_2)\cdots(i_n,t_n)\in(I\times\mathbb{R}_{\geq 0})^*$, a \emph{deterministic timed Mealy machine} (DTMM) $\mathcal{M}$ returns at most one output sequence $\mathcal{M}(\omega)=o_1 o_2 \cdots o_n$. Given two DTMMs $\mathcal{M}_1$ and $\mathcal{M}_2$, for any timed word $\omega$ over inputs $I$, if the output sequences of two DTMMs are equal, i.e., $\mathcal{M}_1(\omega)=\mathcal{M}_2(\omega)$, then the two DTMMs are equivalent, denoted as $\mathcal{M}_1 \approx \mathcal{M}_2$. We modify the learning algorithm to take into account of inputs and outputs. 
By the same argument as for DOTAs, we can show correctness and termination of the learning algorithm for DTMMs.

\begin{theorem}[Correctness and termination for learning DTMMs]
The learning algorithm for DTMMs always terminates and returns a correct DTMM.
\end{theorem}

\section{Implementation and Experiments}
\label{sec:experiment}
To investigate the efficiency and scalability of our methods, we implemented a prototype in Python named $\textsf{SL}$ for both DOTAs and DTMMs based on the tool provided in~\cite{AnCZZZ20}. We use Z3~\cite{MouraB08} as the constraint solving engine. 
We describe some detailed aspects of the implementation below. The implementation and models used for experiments are available at \url{https://github.com/Leslieaj/DOTALearningSMT}. 

\begin{description}
\item[Incremental solving] Our implementation takes advantage of incremental SMT solving functionality in Z3. This allows Z3 to reuse information from previous calls to accelerate the solving process. 
For each query, we push a backtracking point after adding all new constraints in $C_1\wedge C_2\wedge C_4$, then insert the constraints $C_3$ or $C_3'$ depending on the stage of the algorithm. After the query finished, we pop to the previous backtracking point, hence removing $C_3$ or $C_3'$ before the next query.

\item[Sink location] We use a sink location to denote timed behaviors that are invalid or non-described, and it is sometimes possible for membership queries to return whether a timed word reached a sink location. Our implementation takes advantage of this information when it is available. This results in a significant acceleration of the learning process. The technique has been introduced in the previous work~\cite{AnCZZZ20}. 

\item[Equivalence query] Equivalence between timed automata with one clock is decidable. We implemented an equivalence query oracle based on~\cite{OuaknineW04}, but simplified for the deterministic case. In actual applications when the target automaton is unknown, this can be usually replaced by techniques based on conformance testing.
\end{description}

\vspace{-1mm}
We evaluated our prototype tool on two benchmarks that are previously used in~\cite{AnCZZZ20} and~\cite{VaandragerB021}. 
They respectively contain hundreds of randomly generated DOTAs and several models from practical applications which are in the form of DOTAs and MM1Ts. The models from practical applications consist of the abstract automata of an Authentication and Key Management service of the WiFi (AKM), the functional specification of TCP protocol, a car alarm system (CAS), and a
particle counter (PC). All experiments
have been carried out on an Intel Core i7-9750H @ 2.6GHz processor with 16GB RAM running Ubuntu 20.04 Linux system. 

\subsection{Experiments on DOTAs}
\begin{table}[!t]
\caption{Experimental results on DOTAs.}
\label{tab:experiment_DOTAs}
\vspace*{-0.3cm}
\scriptsize
\begin{center}
\resizebox{0.7\textwidth}{!}{
	\begin{tabular}{ccrcrrrrrrrrrr}
		\toprule
		\multirow{2}*{\textbf{Group}} & \multirow{2}*{$\lvert\Delta\rvert$} & & \multirow{2}*{\textbf{Method}} &  \multicolumn{3}{c}{\textbf{{\#}Membership}} & & \multicolumn{3}{c}{\textbf{{\#}Equivalence}} & \multirow{2}*{$\vert Q_{\mathcal{H}}\rvert$} & \multirow{2}*{\textbf{{\#}Learnt}} & \multirow{2}*{$t(s)$}\\
		\cmidrule{5-7} \cmidrule{9-11}
		& & & & $N_{\text{min}}$ & $N_{\text{mean}}$ & $N_{\text{max}}$ & & $N_{\text{min}}$ & $N_{\text{mean}}$ & $N_{\text{max}}$ & & &\\
		\midrule
		\multirow{2}*{6\_2\_10} & \multirow{2}*{11.9} & & \textsf{DOTAL} & 73 & 348.3 & 708 & & 10 & 16.7 & 30 & 5.6 & 7/10 & 39.88 \\
		& & & \textsf{SL} & 104 & 1894.8 & 3929 & & 11 & 20.8 & 35 & 5.6 & 10/10 &0.78 \\
		\midrule[0.05pt]
		\multirow{2}*{4\_4\_20} & \multirow{2}*{16.3} & & \textsf{DOTAL} & 231 & 317.0 & 564 & & 27 & 30.8 & 40 & 4.0 & 6/10 & 100.22 \\
		& & & \textsf{SL} & 1740 & 3497.7 & 5329 & & 24 & 32.8 & 42 & 4.0 & 10/10 &1.42 \\
		\midrule[0.05pt]
		\midrule[0.05pt]
        {7\_4\_20} & {26.0} & &  \textsf{SL} & 6092 & 9393.3 & 15216 & & 44 & 51.5 & 69 & 7.0 & 10/10 &2.90 \\
		\midrule[0.05pt]
		{10\_4\_20} & {39.1} & &  \textsf{SL} & 8579 & 16322.3 & 23726 & & 59 & 76.5 & 93 & 10.0 & 10/10 & 5.89 \\
		\midrule[0.05pt]
        {12\_4\_20} & {47.6} & & \textsf{SL} & 13780 & 20345.5 & 29011 & & 70 & 88.0 & 102 & 12.0 & 10/10 &10.05 \\
		\midrule[0.05pt]
        {14\_4\_20} & {58.4} & &  \textsf{SL} & 18915 & 28569.0 & 40693 & & 92 & 110.6 & 126 & 14.0 & 10/10 &14.69 \\
		\midrule[0.05pt]
        AKM (17\_12\_5) & {40.0} & &  \textsf{SL} & 3453 & 3453.0 & 3453 & & 49 & 49.0 & 49 & 12 & 1/1 &7.19 \\
		\midrule[0.05pt]
		TCP (22\_13\_2)  & {22.0} & &  \textsf{SL} & 4713 & 4713.0 & 4713 & & 32 & 32.0 & 32 & 20 & 1/1 &19.04 \\
		\midrule[0.05pt]
        CAS (14\_10\_27) & {23.0} & &  \textsf{SL} & 4769 & 4769.0 & 4769 & & 18 & 18.0 & 18 & 14 & 1/1 &126.30 \\
		\midrule[0.05pt]
        PC (26\_17\_10) & {42.0} & &  \textsf{SL} & 10854 & 10854.0 & 10854 & & 28 & 28.0 & 28 & 25 & 1/1 &109.01 \\
		\bottomrule
	\end{tabular}

	}
\end{center}
\begin{minipage}{\textwidth}
\scriptsize{
	\textbf{Group}: each group has ID of the form $|Q|\_|\Sigma|\_\kappa$, where $|Q|$ is the number of locations, $|\Sigma|$ is the size of the alphabet, and $\kappa$ is the maximum constant appearing in the clock constraints.
	$\lvert\Delta\rvert$: average number of transitions of a DOTA in the corresponding group.
	\textbf{Method}: \textsf{DOTAL} and \textsf{SL} represent the method in~\cite{AnCZZZ20} and our method respectively. 
	\textbf{{\#}Membership} \& \textbf{{\#}Equivalence}: number of membership and equivalence queries, respectively. $N_{\text{min}}$: minimal, $N_{\text{mean}}$: mean, $N_{\text{max}}$: maximum.
	$|Q_{\mathcal{H}}|$: average number of locations of the learned automata for each group.
	\textbf{{\#}Learnt}: the number of the learnt DOTAs (learnt/total).
	$t$: average wall-clock time in seconds.
}
\end{minipage}
\vspace*{-3mm}
\end{table}
We first compared the performance of our learning algorithm \textsf{SL} with the algorithm \textsf{DOTAL} of~\cite{AnCZZZ20} in the normal teacher setting (see Section~\ref{sec:preliminaries}). In~\cite{AnCZZZ20}, the generated random DOTAs are up to 14 locations, but the algorithm only managed to learn automata with up to 6 locations. The examples are divided into different groups depending on the number of locations, number of actions, and maximum clock value in guards. Each group contains ten automata. Moreover, we tested translations of practical models to DOTA provided in~\cite{VaandragerB021}. The experimental results are shown in Table~\ref{tab:experiment_DOTAs}.

The algorithm \textsf{DOTAL} fails in all of the larger examples due to time and memory limits. Hence, we omit them in the table. In the two groups of smaller examples $6\_2\_10$ and $4\_4\_20$, the algorithm \textsf{DOTAL} can learn some of the cases. In the comparison between number of membership and equivalence queries, we see that \textsf{SL} takes about the same number of equivalence queries, and several times more membership queries. This is likely due to the fact that we exhaustively test all pairs of rows in the table under all reset conditions. However, the algorithm \textsf{SL} is scalable to much larger examples than \textsf{DOTAL}.
\textsf{SL} also successfully learns the DOTA models of four practical applications which are all bigger than the randomly generated DOTAs, and far above the ability of the \textsf{DOTAL} algorithm. This shows the potential of \textsf{SL} in real applications.

\subsection{Experiments on TMMs}

\begin{table}[!t]
\caption{Experimental results on DTMMs and MM1Ts.}
\label{tab:experiment_TMMs}
\vspace*{-0.3cm}
\scriptsize
\begin{center}
\begin{tabular}{cc|rcrcrrrc|rcrcrrr}
	\toprule
	\multirow{2}*{\textbf{Case}}& & \multicolumn{3}{c}{\textbf{DTMM}}& & \multicolumn{3}{c}{\textbf{SL}}& & \multicolumn{3}{c}{\textbf{MM1T}} & & \multicolumn{3}{c}{\textbf{MM1T-$L_{M}^*$}~\cite{VaandragerB021}}  \\
	& & $\lvert Q \rvert$ & $\lvert I \rvert$ & $\lvert \Delta \rvert$ & & \textbf{\#{M}} & \textbf{\#{E}} & $t(s)$ & & $\lvert Q \rvert$ & $\lvert I \rvert$ & $\lvert \Delta \rvert$ & & \textbf{\#{R}} & \textbf{\#{I}} & $t(s)$\\
	\midrule
	AKM & & 
	5 & 5 & 28 & & 691 & 34 & 2.6& & 
	4 & 5 & 24 & & 5361 & 29693 & 5070.4\\
	TCP & & 
	11 & 8 & 19 & & 751 & 10 & 1.9& & 
	11 & 8 & 19 & & 401 & 1868 & 65.7\\
	CAS & & 
	8 & 4 & 17 & & 1654 & 21 & 17.1& & 
	8 & 4 & 17 & & 494 & 2528 & 79.5\\
	AKM & & 
	8 & 8 & 24 & & 1194 & 27 & 6.8& & 
	8 & 8 & 24 & & 392 & 1864 & 85.1\\
	
	\bottomrule
\end{tabular}
\end{center}

\vspace*{-8mm}
\end{table}

We also evaluated our learning algorithm for timed Mealy machines. We first transformed the four MM1T models to DTMMs. As shown in Table~\ref{tab:experiment_TMMs}, for each practical application, its DTMM model is more succinct than the corresponding DOTA model in Table~\ref{tab:experiment_DOTAs}. The size of the DTMM model is also comparable to the size of the MM1T model (the two are equal except the AKM case).

We then run our learning algorithm \textsf{SL} on these models. Compared to learning the corresponding DOTA, learning DTMM takes fewer membership and equivalence queries, except for taking more equivalence queries in the case CAS. Hence, we find DTMMs to be more suitable for learning timed reactive systems than DOTAs. We also run the experiment on MM1T using the algorithm \textbf{MM1T-$L_{M}^*$}. As reported in~\cite{VaandragerB021}, the performance is evaluated according to the total number of resets to the system under learning (SUL) \textbf{\#{R}} and the total number of the performed input actions \textbf{\#{I}}. As these are not directly comparable to number of membership and equivalence queries, we list the results side-by-side in the table. Note also that their implementation is based on LearnLib~\cite{Isberner15} and uses a \emph{random word} equivalence oracle with 1000 tests, while we conduct an exact equivalence checking. The computation time is listed to show that our method can learn the examples efficiently, but the computation times are not directly comparable across methods with different ways to conduct equivalence queries. We also note that~\cite{VaandragerB021} showed experimentally that the algorithm \textbf{MM1T-$L_{M}^*$} compares favorably against heuristic learning methods based on genetic programming~\cite{TapplerALL19,AichernigPT20}.

\section{Conclusion}
\label{sec:conclusion}
In this paper, we proposed a new algorithm for active learning of deterministic one-clock timed automata and timed Mealy machines, using constraint solving based on SMT solving to determine resets and location assignments. This takes advantage of the ability of SMT to solve large constraint systems efficiently, allowing the algorithm to scale up to much larger timed automata models.

In future work, we wish to consider extension of the algorithm to learning timed automata with multiple clocks as well as the non-determinstic case. We wish to also consider incorporating ideas from algorithms such as TTT in order to improve efficiency, in particular reducing the number of membership queries.

%
%
%
\bibliographystyle{splncs04}
\bibliography{paper}
\newpage
\begin{subappendices}
\renewcommand{\thesection}{\Alph{section}} 

\section{Proofs of Theorem~\ref{thm:hypothesis} and Theorem~\ref{thm:correctness}}

\begin{proof}[of Theorem~\ref{thm:hypothesis}]
Since $\overline{\bv_\omega}$ and $\overline{\qv_\omega\vphantom{\bv_\omega}}$ are feasible solutions to the constraints, all constraints in $C_1\wedge C_2\wedge C_3\wedge C_4$ are satisfied. The construction of the hypothesis guarantees that for each transition of the timed word $\omega\in\ssr$, whether reset occurs after the transition agrees with the corresponding assignment of $\mathbb{b}_\omega$. Constraint~\ref{con1} shows that rows in $\ssr$ assigned to the same location must be either all accepting or all non-accepting. Constraint~\ref{con2} shows that the resulting candidate automaton is deterministic, and during the construction of transitions using the partition function (Definition~\ref{def:partition}), no two of the time values $\mu_i$ are in the same region. Constraint~\ref{con3} shows each row in $\ssr$ corresponds to some location. Constraint~\ref{con4} is for efficiency only, the condition that rows in $S$ are assigned to different rows are already enforced by other constraints.

We prove by induction that for each row $\omega\in\ssr$, the location reached by $\omega$ in $\mathcal{H}$ agrees with its assignment $\overline{\mathbb{q}_{\omega}\vphantom{\bv_{\omega}}}$. This is clear for the empty timed word $\epsilon$. Now suppose $\omega$ is of the form $\omega_1\cdot (\sigma_n,t_n)$. Since $\ssr$ is prefix-closed, we have $\omega_1\in\ssr$ and the location reached by $\omega_1$ agrees with $\overline{\mathbb{q}_{\omega_1}\vphantom{\bv_{\omega_1}}}$. Then, according to the construction, the auxiliary transition $(\overline{\mathbb{q}_{\omega_1}\vphantom{\bv_{\omega_1}}}, \sigma, \psi, \overline{\mathbb{b}_{\omega}}, \overline{\mathbb{q}_{\omega}\vphantom{\bv_{\omega}}})$ with $\psi$ equal to the clock valuation after executing $\omega$. Then, according to the partition function, a transition will be formed with guard containing $\psi$ from $\overline{\mathbb{q}_{\omega_1}\vphantom{\bv_{\omega_1}}}$ to $\overline{\mathbb{q}_{\omega}\vphantom{\bv_{\omega}}}$ and with action $\sigma$. This shows the location reached by $\omega$ agrees with $\overline{\mathbb{q}_{\omega}\vphantom{\bv_{\omega}}}$.

Next, the fact that each row $\omega\in\ssr$ is accepted by the hypothesis $\mathcal{H}$ iff $\textsf{MQ}(\omega)=+$ follows from the above statement, together with the definition of $Q_{\mathcal{H}}$ and $F_\mathcal{H}$. Finally, for any two rows $\omega_1,\omega_2\in\ssr$, if the value of $f$ on $\omega_1,\omega_2$ and the setting of reset variables $\overline{\mathbb{b}}$ is $\bot$, then $\overline{\qv_{\omega_1}\vphantom{\bv_{\omega_1}}}\neq \overline{\qv_{\omega_2}\vphantom{\bv_{\omega_2}}}$ follows from Constraint~\ref{con1}, and by the statement proved above, they also reach different locations in $\mathcal{H}$. \qed
\end{proof}

\begin{proof}[of Theorem~\ref{thm:correctness}]
Termination of the algorithm is proved by a simulation argument, comparing against a brute-force search version of the algorithm similar to the normal teacher case in~\cite{AnCZZZ20}. The brute-force search constructs a binary tree of observation tables, with each branching corresponds to a choice of reset for some row in $S\cup R$ (since the reset information is now certain, the observation tables are the usual ones for $L^*$ algorithm, and the category $S_+$ is not needed). The search is performed in the breadth-first manner, with respect to the number of rows in $S$. Along the branch where all choice of resets agree with the target automaton, one correct automaton will be eventually learned. This is because different rows in $S$ must arrive at different symbolic states (a pair of location and region) in the automaton, so the size of $S$ is bounded by the number of symbolic states (number of locations multiplied by the number of regions). In practice, the use of partition function means the correct automaton is found when $S$ is much smaller than the number of symbolic states. Along other branches, a correct automaton may or may not be obtained. In the latter case the number of rows in $S$ increase indefinitely. Hence, the breadth-first search must terminate with learning a correct automaton.

Our algorithm simulates the breadth-first search, with the value of $N$ in the observation table corresponding to the current size of $S$. Rows in $S$ in our algorithm corresponds to rows that appear in $S$ along all branches of the search tree, and rows in $S_+$ corresponds to rows that appear in $S$ along some, but not all branches of the search tree. Each model generated by constraint solving corresponds to considering some branch of the tree, but with counterexamples added to all branches at the same time. This shows our algorithm terminates with learning a correct automaton as well. \qed
\end{proof}

\section{Detailed Running Example}
\vspace{-0.2cm}
The following Fig.~\ref{fig:detailed_steps} illustrates the detailed learning steps for the DOTA $\mathcal{A}$ in Fig.~\ref{fig:COTA}. In Step 1, the observation table $\mathcal{O}_1=(S,S_+,R,E,f,N)$ is initialized with $S=E=\{\epsilon\}, S_+=\emptyset, R=\{(a,0)\}, N=1$, then $\mathsf{move\_to\_S}$ operation will move $(a,0)$ to $S$ and add $(a,0)(a,0)$ to $R$ since $(a,0)$ is distinct from $\epsilon$ under any reset. Since $(a,0)(a,0)$ cannot be distinguished from $(a,0)\in S$, the first observation table $\mathcal{O}_1$ is prepared to encode constraints $Cons_1 = C_1\wedge C_2 \wedge C_3 \wedge C_4$. By solving it with the SMT solver, we obtain a model $M$ of the reset variables and location variables. After the hypothesis construction, the DOTA $\mathcal{H}_1$ is built from $\mathcal{O}_1$ and $M$. By performing an equivalence query for $\mathcal{H}_1$, the teacher returns a counterexample $\mathit{ctx}_1=(a,4)$. The counterexample is added to $R$ since it is not distinct from $\epsilon$. $O_2$ is prepared now. After the same process, we get the hypothesis $\mathcal{H}_2$ shown in the Step 2. 
The teacher returns a counterexample $(a,4)(a,5.5)$ and we add it in $R$ after Step 3. However, the inconsistency is find that $(a,4)(a,5.5)$ and $(a,9.5)$ end in the same region under $\bv_0=\bv_3=\bv_5=\bot$, but the timed words $(a,4)$ and $\epsilon$ can not be distinguished with suffix $e=\epsilon$. Then the timed word $(a, \mathsf{min}(5.5,9.5))\cdot e$, i.e., $(a,5.5)$ is added to $E$.
The situation in Example~\ref{expl:algorithm} occurs at Step 4, and as introduced in the example we can obtain the table $\mathcal{O}_5$ in Step 5 by moving $(a,4)(a,5.5)$  to $S_+$. After equivalence query, teacher returns counterexample $(a,4)(a,0)$. Two timed words $(a,0)$ and $(a,0)(a,5.5)$ are added to $E$ during constructing the consistency constraint. Now timed word $(a,4)\in R$ is certainly distinct from all the rows in $S$, which violates Constraint~\ref{con3}, an thus the corresponding constraint $\mathit{Cons}_6$ built from $\mathcal{O}_6$ is unsatisfiable. Therefore, $(a,4)$ is moved to $S$ and a correct automaton $\mathcal{H}_7$ is learned in the last step.

\begin{figure}[h]
\vspace{-0.5cm}
\centering
\begin{subfigure}{\textwidth}
\begin{minipage}{0.5\linewidth}
\centering
    \resizebox{0.8\linewidth}{!}{
    \begin{tabular}{lccc|c|c|c c||c}
		   \toprule
		   \multicolumn{3}{c}{\multirow{2}{*}{$\mathcal{O}_1$}}& &\multicolumn{3}{c}{\textcolor{purple}{$S\cup S_+ \cup R$}} & & \multirow{2}*{\textcolor{purple}{$E$}}\\
		   & & & & $\epsilon$ & $\left(a,0\right)$ & $\left(a,0\right)\left(a, 0\right)$ & & \\
		   \cline{1-8}
		   \multirow{2}*{\textcolor{purple}{$S$}} & & $\epsilon$ & $\mathbb{b}_0$ & $\top$ & $\bot$ & $\bot$ & & $\epsilon$\\
		    & & $(a, 0)$ & $\mathbb{b}_1$ & $\bot$ & $\top$ & $\top$ & \\
		   \cline{1-4}
		   \textcolor{purple}{$R$} & & $(a,0)(a,0)$ & $\mathbb{b}_2$ & $\bot$ & $\top$ & $\top$ & & \\
		  \bottomrule
		\end{tabular}
		}
\end{minipage}
\hfill
\begin{minipage}{0.49\linewidth}
\centering
     \resizebox{0.7\linewidth}{!}{
        \begin{tikzpicture}[->,
			>=stealth,
			node distance = 2.5cm,
			line width=0.4mm,
			every state/.style={thick, fill=gray!10, minimum size=2pt},
			sink/.style={thick, fill=blue!30, minimum size=2pt},
			sink edge/.style={thick, fill=blue!30}
			]
			\node[state, accepting, initial, initial where=left] (q0) {$q_0$};
			\node[state, right = 2.5cm of q0] (q1) {$q_1$};
			\draw (q0) edge[above, black] node{$a, \left[0, +\right), \bot$} (q1);
			\draw (q1) edge[loop above] node{$a, \left[0, +\right), \bot$} (q1);
			\node [below= 1cm of $(q0)!0.5!(q1)$] {$\mathcal{H}_1$};
		\end{tikzpicture}
 		}
\end{minipage}
\vspace{-0.1cm}
\subcaption{Step 1}
\end{subfigure}

\begin{minipage}{1\linewidth}
\centering
\begin{tikzpicture}
        \node[draw, single arrow,
              minimum height=10mm, minimum width=3mm,
              single arrow head extend=2mm,
              anchor=west, rotate=-90] at (8,-1) {};
        \node at (10.5, -1.5) {counterexample: $(a, 4)$};
    \end{tikzpicture}
\end{minipage}

\begin{subfigure}{\textwidth}
\begin{minipage}{0.6\linewidth}
\centering
    \resizebox{0.8\linewidth}{!}{
    \begin{tabular}{lccc|c|c|c|c c||c}
		   \toprule
		   \multicolumn{3}{c}{\multirow{2}{*}{$\mathcal{O}_2$}}& &\multicolumn{4}{c}{\textcolor{purple}{$S\cup S_+ \cup R$}} & & \multirow{2}*{\textcolor{purple}{$E$}}\\
		   & & & & $\epsilon$ & $\left(a,0\right)$ & $\left(a,0\right)\left(a, 0\right)$ & $(a,4)$ & & \\
		   \cline{1-8}
		   \multirow{2}*{\textcolor{purple}{$S$}} & & $\epsilon$ & $\mathbb{b}_0$ & $\top$ & $\bot$ & $\bot$ & $\top$ & & $\epsilon$\\
		    & & $(a, 0)$ & $\mathbb{b}_1$ & $\bot$ & $\top$ & $\top$ & $\bot$ & \\
		   \cline{1-4}
		   \multirow{2}*{\textcolor{purple}{$R$}} & & $(a,0)(a,0)$ & $\mathbb{b}_2$ & $\bot$ & $\top$ & $\top$ & $\bot$ & & \\
		    & & $(a,4)$ & $\mathbb{b}_3$ & $\top$ & $\bot$ & $\bot$ & $\top$ & & \\
		   
		  \bottomrule
		\end{tabular}
		}
\end{minipage}
\hfill
\begin{minipage}{0.39\linewidth}
\centering
     \resizebox{0.9\linewidth}{!}{
        \begin{tikzpicture}[->,
			>=stealth,
			node distance = 2.5cm,
			line width=0.4mm,
			every state/.style={thick, fill=gray!10, minimum size=2pt},
			sink/.style={thick, fill=blue!30, minimum size=2pt},
			sink edge/.style={thick, fill=blue!30}
			]
			\node[state, accepting, initial, initial where=left] (q0) {$q_0$};
			\node[state, right = 2.5cm of q0] (q1) {$q_1$};
			\draw (q0) edge[above, black] node{$a, \left[0, 4\right), \bot$} (q1);
			\draw (q0) edge[loop above] node{$a, \left[4, +\right), \bot$} (q0);
			\draw (q1) edge[loop above] node{$a, \left[0, +\right), \bot$} (q1);
			\node [below= 1cm of $(q0)!0.5!(q1)$] {$\mathcal{H}_2$};
		\end{tikzpicture}
 		}
\end{minipage}
\subcaption{Step 2}
\end{subfigure}
\vspace{-0.3cm}
\caption{The detailed learning steps for the example DOTA $\mathcal{A}$ in Fig.~\ref{fig:COTA}}
\end{figure}
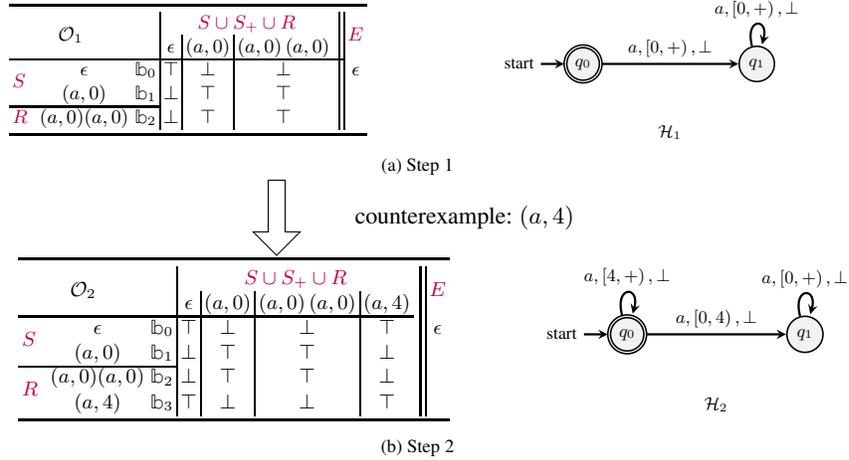

\begin{figure}
\ContinuedFloat
\begin{minipage}{1\linewidth}
\centering
\begin{tikzpicture}
        \node[draw, single arrow,
              minimum height=10mm, minimum width=3mm,
              single arrow head extend=2mm,
              anchor=west, rotate=-90] at (8,-1) {};
        \node at (10.5, -1.5) {counterexample: $(a,9.5)$};
    \end{tikzpicture}
\end{minipage}
 \vspace{0.3cm}
 
\begin{subfigure}{\textwidth}
\begin{minipage}{0.64\linewidth}
\centering
\resizebox{0.8\linewidth}{!}{
    \begin{tabular}{lccc|c|c|c|c|c c||c}
		   \toprule
		   \multicolumn{3}{c}{\multirow{2}{*}{$\mathcal{O}_3$}}& &\multicolumn{5}{c}{\textcolor{purple}{$S\cup S_+ \cup R$}} & & \multirow{2}*{\textcolor{purple}{$E$}}\\
		   & & & & $\epsilon$ & $\left(a,0\right)$ & $\left(a,0\right)\left(a, 0\right)$ & $(a,4)$ & $(a,9.5)$ & & \\
		   \cline{1-9}
		   \multirow{2}*{\textcolor{purple}{$S$}} & & $\epsilon$ & $\mathbb{b}_0$ & $\top$ & $\bot$ & $\bot$ & $\top$ & $\bot$ & & $\epsilon$\\
		    & & $(a, 0)$ & $\mathbb{b}_1$ & $\bot$ & $\top$ & $\top$ & $\bot$ &$\top$ &\\
		   \cline{1-4}
		   \multirow{3}*{\textcolor{purple}{$R$}} & & $(a,0)(a,0)$ & $\mathbb{b}_2$ & $\bot$ & $\top$ & $\top$ & $\bot$ & $\top$ & & \\
		    & & $(a,4)$ & $\mathbb{b}_3$ & $\top$ & $\bot$ & $\bot$ & $\top$ & $\bot$ & & \\
		   & & $(a,9.5)$ & $\mathbb{b}_4$ & $\bot$ & $\top$ & $\top$ & $\bot$ & $\top$ & &\\
		  \bottomrule
		\end{tabular}
		}
\end{minipage}
\hfill
\begin{minipage}{0.35\linewidth}
\centering
    \resizebox{0.95\linewidth}{!}{
        \begin{tikzpicture}[->,
			>=stealth,
			node distance = 2.5cm,
			line width=0.4mm,
			every state/.style={thick, fill=gray!10, minimum size=2pt},
			sink/.style={thick, fill=blue!30, minimum size=2pt},
			sink edge/.style={thick, fill=blue!30}
			]
			\node[state, accepting, initial, initial where=left] (q0) {$q_0$};
			\node[state, right = 2.2cm of q0] (q1) {$q_1$};
			\draw (q0) edge[bend left, above, black] node{$a, \left[0, 4\right), \bot$} (q1);
			\draw (q0) edge[bend right, below, black] node{$a, \left(9, +\right), \bot$} (q1);
			\draw (q0) edge[loop above] node{$a, \left[4, 9\right], \bot$} (q0);
			\draw (q1) edge[loop above] node{$a, \left[0, +\right), \bot$} (q1);
			\node [below= 1.7cm of $(q0)!0.5!(q1)$] {$\mathcal{H}_3$};
		\end{tikzpicture}
		}
\end{minipage}
\caption{Step 3}
\end{subfigure}
\vspace{0.3cm}

\begin{minipage}{1\linewidth}
\centering
\begin{tikzpicture}
        \node[draw, single arrow,
              minimum height=10mm, minimum width=3mm,
              single arrow head extend=2mm,
              anchor=west, rotate=-90] at (11,-1) {};
        \node at (13.3, -1.5) {counterexample: $(a,4)(a,5.5)$};
    \end{tikzpicture}
\vspace{0.3cm}
\end{minipage}

\begin{subfigure}{\textwidth}

\begin{minipage}{0.8\linewidth}
\centering
    \resizebox{0.95\linewidth}{!}{
    \begin{tabular}{lccc|c|c|c|c|c|c c||c}
		   \toprule
		   \multicolumn{3}{c}{\multirow{2}{*}{$\mathcal{O}_4$}}& &\multicolumn{6}{c}{\textcolor{purple}{$S\cup S_+ \cup R$}} & & \multirow{2}*{\textcolor{purple}{$E$}}\\
		   & & & & $\epsilon$ &$\left(a, 0\right)$ & $\left(a,4\right)\left(a,5.5\right)$ & $\left(a,0\right)\left(a, 0\right)$ & $\left(a,4\right)$ & $\left(a, 9.5\right)$ & & \\
		   \cline{1-10}
		   \textcolor{purple}{$S$} & & $\epsilon$ & \multicolumn{1}{c|}{$\mathbb{b}_0$} & \multicolumn{1}{c|}{$\top$} & \multicolumn{1}{c|}{$\bot$} & \multicolumn{1}{c|}{$\neg \mathbb{b}_5$} & \multicolumn{1}{c|}{$\bot$} & \multicolumn{1}{c|}{$\mathbb{b}_3$} & $\bot$ & & \multirow{2}*{$\epsilon$}\\
		   & & $(a, 0)$ & \multicolumn{1}{c|}{$\mathbb{b}_1$} &\multicolumn{1}{c|}{$\bot$} & \multicolumn{1}{c|}{$\top$} & \multicolumn{1}{c|}{$\bot$} & \multicolumn{1}{c|}{$\top$} & \multicolumn{1}{c|}{$\bot$} & $\top$ & & \\
		   \cline{1-4}
            \multirow{4}*{\textcolor{purple}{$R$}} & & $\left(a, 4\right)\left(a,5.5\right)$ & $\mathbb{b}_5$  & \multicolumn{1}{c|}{$\neg \mathbb{b}_5$} & \multicolumn{1}{c|}{$\bot$} & \multicolumn{1}{c|}{$\top$} & \multicolumn{1}{c|}{$\bot$} & \multicolumn{1}{c|}{$\bot$} & $\bot$ & & $\left(a,5.5\right)$ \\
		    & & $\left(a, 0\right)\left(a, 0\right)$ & $\mathbb{b}_2$ & \multicolumn{1}{c|}{$\bot$} & \multicolumn{1}{c|}{$\top$} & \multicolumn{1}{c|}{$\bot$} & \multicolumn{1}{c|}{$\top$} & \multicolumn{1}{c|}{$\bot$} & $\top$ & & \\
		   & & $\left(a, 4\right)$ & $\mathbb{b}_3$  & \multicolumn{1}{c|}{$\mathbb{b}_3$} & \multicolumn{1}{c|}{$\bot$} & \multicolumn{1}{c|}{$\mathbf{\bot}$} & \multicolumn{1}{c|}{$\bot$} & \multicolumn{1}{c|}{$\top$} & $\bot$ & & \\
		   & & $\left(a, 9.5\right)$ & $\mathbb{b}_4$  & \multicolumn{1}{c|}{$\bot$} & \multicolumn{1}{c|}{$\top$} & \multicolumn{1}{c|}{$\bot$} & \multicolumn{1}{c|}{$\top$} & \multicolumn{1}{c|}{$\bot$} & $\top$ & & \\
		  \bottomrule
		\end{tabular}
	}
\end{minipage}
\hfill
\begin{minipage}{0.1\linewidth}
\textbf{UNSAT}
\end{minipage}
\subcaption{Step 4}
\end{subfigure}

\begin{minipage}{1\linewidth}
\centering
\begin{tikzpicture}
        \node[draw, single arrow,
              minimum height=10mm, minimum width=3mm,
              single arrow head extend=2mm,
              anchor=west, rotate=-90] at (8,-1) {};
        \node at (10.5, -1.5) {Move $(a,4)(a,5.5)$ to $S_+$};
    \end{tikzpicture}
\end{minipage}

\begin{subfigure}{\textwidth}
\hspace{-0.5cm}
\begin{minipage}{0.75\linewidth}
\centering
    \resizebox{\linewidth}{!}{
    \begin{tabular}{lccc|c|c|c|c|c|c c||c}
		   \toprule
		   \multicolumn{3}{c}{\multirow{2}{*}{$\mathcal{O}_5$}} & &\multicolumn{6}{c}{\textcolor{purple}{$S\cup S_+ \cup R$}} & & \multirow{2}*{\textcolor{purple}{$E$}}\\
		   & & & & $\epsilon$ &$\left(a, 0\right)$ & $\left(a,4\right)\left(a,5.5\right)$ & $\left(a,0\right)\left(a, 0\right)$ & $\left(a,4\right)$ & $\left(a, 9.5\right)$ & & \\
		   \cline{1-10}
		   \textcolor{purple}{$S$} & & $\epsilon$ & \multicolumn{1}{c|}{$\mathbb{b}_0$} & \multicolumn{1}{c|}{$\top$} & \multicolumn{1}{c|}{$\bot$} & \multicolumn{1}{c|}{$\neg \mathbb{b}_5$} & \multicolumn{1}{c|}{$\bot$} & \multicolumn{1}{c|}{$\mathbb{b}_3$} & $\bot$ & & \multirow{2}*{$\epsilon$}\\
		   & & $(a, 0)$ & \multicolumn{1}{c|}{$\mathbb{b}_1$} &\multicolumn{1}{c|}{$\bot$} & \multicolumn{1}{c|}{$\top$} & \multicolumn{1}{c|}{$\bot$} & \multicolumn{1}{c|}{$\top$} & \multicolumn{1}{c|}{$\bot$} & $\top$ & & \\
		   \cline{1-4}
            \textcolor{purple}{$S_+$} & & $\left(a, 4\right)\left(a,5.5\right)$ & $\mathbb{b}_5$  & \multicolumn{1}{c|}{$\neg \mathbb{b}_5$} & \multicolumn{1}{c|}{$\bot$} & \multicolumn{1}{c|}{$\top$} & \multicolumn{1}{c|}{$\bot$} & \multicolumn{1}{c|}{$\bot$} & $\bot$ & & $\left(a,5.5\right)$\\
		   \cline{1-4}
		   \multirow{3}*{\textcolor{purple}{$R$}} & & $\left(a, 0\right)\left(a, 0\right)$ & $\mathbb{b}_2$ & \multicolumn{1}{c|}{$\bot$} & \multicolumn{1}{c|}{$\top$} & \multicolumn{1}{c|}{$\bot$} & \multicolumn{1}{c|}{$\top$} & \multicolumn{1}{c|}{$\bot$} & $\top$ & & \\
		   & & $\left(a, 4\right)$ & $\mathbb{b}_3$  & \multicolumn{1}{c|}{$\mathbb{b}_3$} & \multicolumn{1}{c|}{$\bot$} & \multicolumn{1}{c|}{$\mathbf{\bot}$} & \multicolumn{1}{c|}{$\bot$} & \multicolumn{1}{c|}{$\top$} & $\bot$ & & \\
		   & & $\left(a, 9.5\right)$ & $\mathbb{b}_4$  & \multicolumn{1}{c|}{$\bot$} & \multicolumn{1}{c|}{$\top$} & \multicolumn{1}{c|}{$\bot$} & \multicolumn{1}{c|}{$\top$} & \multicolumn{1}{c|}{$\bot$} & $\top$ & & \\
		  \bottomrule
		\end{tabular}
	}
\end{minipage}
\hfill
\begin{minipage}{0.24\linewidth}
\centering
    \resizebox{1.3\linewidth}{!}{
        \begin{tikzpicture}[->,
			>=stealth,
			node distance = 2.5cm,
			line width=0.4mm,
			every state/.style={thick, fill=gray!10, minimum size=2pt},
			sink/.style={thick, fill=blue!30, minimum size=2pt},
			sink edge/.style={thick, fill=blue!30}
			]
			\node[state, accepting, initial, initial where=left] (q0) {$q_0$};
			\node[state, accepting, right = 2.5cm of q0] (q2) {$q_2$};
			\node[state, below= 1.8cm of $(q0)!0.5!(q2)$] (q1) {$q_1$};
			\draw (q0) edge[loop above] node{$a, \left[4, 5\right], \top$} (q0);
			\draw (q0) edge[above, black] node{$a, \left(5, 9\right], \bot$} (q2);
			\draw (q0) edge[sloped, anchor=center, below, line width=0.4mm] node{$a, \left[0, 4\right). \bot$} (q1);
			\draw (q2) edge[sloped, anchor=center, below] node{$a, \left(5, +\right), \bot$} (q1);
			\draw (q1) edge[loop below, line width=0.4mm] node{$a, \left[0, +\right), \bot$} (q1);
			\draw (q0) edge[sloped, below, bend right=50] node{$a, \left(9, \infty\right), \top$} (q1);
			\node [below= 3.4cm of $(q0)!0.5!(q2)$] {$\mathcal{H}_5$};
		\end{tikzpicture}
		}
\end{minipage}
\subcaption{Step 5}
\end{subfigure}
\caption{The detailed learning steps for the example DOTA $\mathcal{A}$ in Fig.~\ref{fig:COTA} (Continued).}
\end{figure}

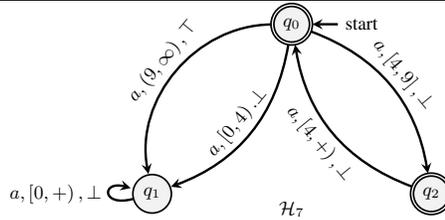
\begin{figure}[!t]
\ContinuedFloat

\begin{minipage}{1\textwidth}
\centering
\begin{tikzpicture}
        \node[draw, single arrow,
              minimum height=10mm, minimum width=3mm,
              single arrow head extend=2mm,
              anchor=west, rotate=-90] at (8,-1) {};
        \node at (10.5, -1.5) {Counterexample: $(a,4)(a,0)$};
    \end{tikzpicture}
\end{minipage}
\vspace{0.3cm}

\begin{subfigure}{\linewidth}
\hspace{-2cm}
\begin{minipage}{\linewidth}
\centering
    \resizebox{1.3\linewidth}{!}{
    \begin{tabular}{lccc|c|c|c|c|c|c|c|c c||c}
		   \toprule
		   \multicolumn{3}{c}{\multirow{2}{*}{$\mathcal{O}_6$}}& &\multicolumn{8}{c}{\textcolor{purple}{$S\cup S_+ \cup R$}} & & \multirow{2}*{\textcolor{purple}{$E$}}\\
		   & & & & $\epsilon$ &$\left(a, 0\right)$ & $\left(a,4\right)\left(a,5.5\right)$ & $\left(a,0\right)\left(a, 0\right)$ & $\left(a,4\right)$ & $\left(a, 9.5\right)$ & $\left(a,4\right)\left(a,5.5\right)\left(a,0\right)$ & $\left(a,4\right)\left(a,0\right)$ & & \\
		   \cline{1-12}
		   \textcolor{purple}{$S$} & & $\epsilon$ & \multicolumn{1}{c|}{$\mathbb{b}_0$} & \multicolumn{1}{c|}{$\top$} & \multicolumn{1}{c|}{$\bot$} & \multicolumn{1}{c|}{$\neg \mathbb{b}_5 \wedge \neg(\mathbb{b}_3\wedge \neg \mathbb{b}_5)$} & \multicolumn{1}{c|}{$\bot$} & \multicolumn{1}{c|}{$\bot$} & $\bot$ & $\bot$ & $\neg (\mathbb{b}_3\wedge \neg \mathbb{b}_7) \wedge \neg \mathbb{b}_7$ & & \multirow{2}*{$\epsilon$}\\
		   & & $(a, 0)$ & \multicolumn{1}{c|}{$\mathbb{b}_1$} &\multicolumn{1}{c|}{$\bot$} & \multicolumn{1}{c|}{$\top$} & \multicolumn{1}{c|}{$\bot$} & \multicolumn{1}{c|}{$\top$} & \multicolumn{1}{c|}{$\bot$} & $\top$ & $\top$ & $\bot$ & & \\
		   \cline{1-4}
            \textcolor{purple}{$S_+$} & & $\left(a, 4\right)\left(a,5.5\right)$ & $\mathbb{b}_5$  & \multicolumn{1}{c|}{$\neg \mathbb{b}_5 \wedge \neg (\mathbb{b}_3\wedge \neg \mathbb{b}_5)$} & \multicolumn{1}{c|}{$\bot$} & \multicolumn{1}{c|}{$\top$} & \multicolumn{1}{c|}{$\bot$} & \multicolumn{1}{c|}{$\bot$} & $\bot$ & $\bot$ & $\neg \mathbb{b}_5$ & & \multirow{2}*{$(a, 5.5)$}\\
		   \cline{1-4}
		   \multirow{5}*{\textcolor{purple}{$R$}} & & $\left(a, 0\right)\left(a, 0\right)$ & $\mathbb{b}_2$ & \multicolumn{1}{c|}{$\bot$} & \multicolumn{1}{c|}{$\top$} & \multicolumn{1}{c|}{$\bot$} & \multicolumn{1}{c|}{$\top$} & \multicolumn{1}{c|}{$\bot$} & $\top$ & $\top$ & $\bot$ & & \\
		   & & $\left(a, 4\right)$ & $\mathbb{b}_3$  & \multicolumn{1}{c|}{$\bot$} & \multicolumn{1}{c|}{$\bot$} & \multicolumn{1}{c|}{$\mathbf{\bot}$} & \multicolumn{1}{c|}{$\bot$} & \multicolumn{1}{c|}{$\top$} & $\bot$ & $\bot$ & $\bot$ & & \multirow{2}*{$\left(a,0\right)$}\\
		   & & $\left(a, 9.5\right)$ & $\mathbb{b}_4$  & \multicolumn{1}{c|}{$\bot$} & \multicolumn{1}{c|}{$\top$} & \multicolumn{1}{c|}{$\bot$} & \multicolumn{1}{c|}{$\top$} & \multicolumn{1}{c|}{$\bot$} & $\top$ & $\top$ & $\bot$& & \\
		   & & $\left(a, 4\right)\left(a,5.5\right)\left(a,0\right)$ & $\mathbb{b}_6$  & \multicolumn{1}{c|}{$\bot$} & \multicolumn{1}{c|}{$\top$} & \multicolumn{1}{c|}{$\bot$} & \multicolumn{1}{c|}{$\top$} & \multicolumn{1}{c|}{$\bot$} & $\top$ & $\top$ & $\bot$ & & \multirow{2}*{$\left(a,0\right)\left(a,5.5\right)$}\\
		   & & $\left(a, 4\right)\left(a,0\right)$ & $\mathbb{b}_7$  & \multicolumn{1}{c|}{$\neg (\mathbb{b}_3\wedge \neg \mathbb{b}_7) \wedge \neg \mathbb{b}_7$} & \multicolumn{1}{c|}{$\top$} & \multicolumn{1}{c|}{$\neg \mathbb{b}_5$} & \multicolumn{1}{c|}{$\top$} & \multicolumn{1}{c|}{$\bot$} & $\bot$ & $\bot$ & $\top$ & & \\
		  \bottomrule
		\end{tabular}
 	}
\end{minipage}

\begin{minipage}{\linewidth}
\hspace{5cm}
\begin{center}
    \textbf{UNSAT}
\end{center}
\end{minipage}
\subcaption{Step 6}
\end{subfigure}

\begin{minipage}{1\textwidth}
\centering
\begin{tikzpicture}
        \node[draw, single arrow,
              minimum height=10mm, minimum width=3mm,
              single arrow head extend=2mm,
              anchor=west, rotate=-90] at (8,-1) {};
        \node at (10.5, -1.5) {Move $(a,4)$ to $S$};
    \end{tikzpicture}
\end{minipage}
\vspace{0.3cm}

\begin{subfigure}{\textwidth}
\hspace{-2cm}
\begin{minipage}{\linewidth}
\centering
    \resizebox{1.3\linewidth}{!}{
    \begin{tabular}{lccc|c|c|c|c|c|c|c|c c||c}
		   \toprule
		   \multicolumn{3}{c}{\multirow{2}{*}{$\mathcal{O}_7$}}& &\multicolumn{8}{c}{\textcolor{purple}{$S\cup S_+ \cup R$}} & & \multirow{2}*{\textcolor{purple}{$E$}}\\
		   
		   & & & & $\epsilon$ & 
		   $\left(a, 0\right)$ & 
		   $\left(a,4\right)$ &
		   $\left(a,4\right)\left(a,5.5\right)$ & 
		   $\left(a,0\right)\left(a, 0\right)$ & 
		   $\left(a, 9.5\right)$ & $\left(a,4\right)\left(a,5.5\right)\left(a,0\right)$ & $\left(a,4\right)\left(a,0\right)$ & & \\
		   \cline{1-12}
		   
		   \multirow{3}*{\textcolor{purple}{$S$}} & & $\epsilon$ & \multicolumn{1}{c|}{$\mathbb{b}_0$} & 
		   \multicolumn{1}{c|}{$\top$} & 
		   \multicolumn{1}{c|}{$\bot$} & 
		   \multicolumn{1}{c|}{$\bot$} &
		   \multicolumn{1}{c|}{$\neg \mathbb{b}_5 \wedge \neg (\mathbb{b}_3\wedge \neg \mathbb{b}_5)$} & 
		   \multicolumn{1}{c|}{$\bot$} & 
		   $\bot$ & $\bot$ & $\neg (\mathbb{b}_3\wedge \neg \mathbb{b}_7) \wedge \neg \mathbb{b}_7$ 
		   & & \multirow{2}*{$\epsilon$}\\
		   
		   & & $(a, 0)$ & \multicolumn{1}{c|}{$\mathbb{b}_1$} &
		   \multicolumn{1}{c|}{$\bot$} & 
		   \multicolumn{1}{c|}{$\top$} & 
		   \multicolumn{1}{c|}{$\bot$} &
		   \multicolumn{1}{c|}{$\bot$} & 
		   \multicolumn{1}{c|}{$\top$} & 
		   $\top$ &
		   $\top$ & 
		   $\bot$ & & \\
		   
		   & & $\left(a, 4\right)$ & $\mathbb{b}_3$  & 
		   \multicolumn{1}{c|}{$\bot$} & 
		   \multicolumn{1}{c|}{$\bot$} & 
		   \multicolumn{1}{c|}{$\top$} &
		   \multicolumn{1}{c|}{$\mathbf{\bot}$} & 
		   \multicolumn{1}{c|}{$\bot$} & 
		   $\bot$ & $\bot$ & $\bot$ & & \multirow{2}*{$(a, 5.5)$}\\
		   \cline{1-4}
            \textcolor{purple}{$S_+$} & & $\left(a, 4\right)\left(a,5.5\right)$ & $\mathbb{b}_5$  & 
            \multicolumn{1}{c|}{$\neg \mathbb{b}_5 \wedge \neg(\mathbb{b}_3\wedge \neg \mathbb{b}_5)$} & 
            \multicolumn{1}{c|}{$\bot$} & 
            \multicolumn{1}{c|}{$\bot$} & 
            \multicolumn{1}{c|}{$\top$} & 
            \multicolumn{1}{c|}{$\bot$} & 
            $\bot$ & $\bot$ & $\neg \mathbb{b}_5$ & & \\
		   \cline{1-4}
		   
		   \multirow{5}*{\textcolor{purple}{$R$}} & & $\left(a, 0\right)\left(a, 0\right)$ & $\mathbb{b}_2$ & 
		   \multicolumn{1}{c|}{$\bot$} & 
		   \multicolumn{1}{c|}{$\top$} & 
		   \multicolumn{1}{c|}{$\bot$} & 
		   \multicolumn{1}{c|}{$\bot$} & 
		   \multicolumn{1}{c|}{$\top$} & 
		   $\top$ & $\top$ & $\bot$ & & \multirow{2}*{$\left(a,0\right)$}\\
		   
		   & & $\left(a, 9.5\right)$ & $\mathbb{b}_4$  & 
		   \multicolumn{1}{c|}{$\bot$} & 
		   \multicolumn{1}{c|}{$\top$} & 
		   \multicolumn{1}{c|}{$\bot$} &
		   \multicolumn{1}{c|}{$\bot$} & 
		   \multicolumn{1}{c|}{$\top$} & 
		   $\top$ & $\top$ & $\bot$& & \\
		   
		   & & $\left(a, 4\right)\left(a,5.5\right)\left(a,0\right)$ & $\mathbb{b}_6$  & 
		   \multicolumn{1}{c|}{$\bot$} & 
		   \multicolumn{1}{c|}{$\top$} & 
		   \multicolumn{1}{c|}{$\bot$} & 
		   \multicolumn{1}{c|}{$\bot$} & 
		   \multicolumn{1}{c|}{$\top$} & 
		   $\top$ & $\top$ & $\bot$ & & \multirow{2}*{$\left(a,0\right)\left(a,5.5\right)$}\\
		   
		   & & $\left(a, 4\right)\left(a,0\right)$ & $\mathbb{b}_7$  & \multicolumn{1}{c|}{$\neg (\mathbb{b}_3\wedge \neg \mathbb{b}_7) \wedge \neg\mathbb{b}_7$} & 
		   \multicolumn{1}{c|}{$\top$} & 
		   \multicolumn{1}{c|}{$\bot$} & 
		   \multicolumn{1}{c|}{$\neg \mathbb{b}_5$} & 
		   \multicolumn{1}{c|}{$\top$} & 
		   $\bot$ & 
		   $\bot$ & $\top$ & & \\
		  \bottomrule
		\end{tabular}
 	}
\end{minipage}

\begin{minipage}{\linewidth}
\hspace{2.5cm}
    \resizebox{0.5\linewidth}{!}{
        \begin{tikzpicture}[->,
			>=stealth,
			node distance = 2.5cm,
			line width=0.4mm,
			every state/.style={thick, fill=gray!10, minimum size=2pt},
			sink/.style={thick, fill=blue!30, minimum size=2pt},
			sink edge/.style={thick, fill=blue!30}
			]
            \node[state] (q1) {$q_1$};
            \node[state, accepting, right = 4cm of q1] (q2) {$q_2$};
            \node[state, accepting, initial, above= 2.5cm of $(q1)!0.5!(q2)$, initial where=right] (q0) {$q_0$};
			\draw (q0) edge[sloped, anchor=center, bend left, above, black] node{$a, \left[4, 9\right], \bot$} (q2);
			\draw (q0) edge[sloped, anchor=center, bend left, above, line width=0.4mm] node{$a, \left[0, 4\right). \bot$} (q1);
			\draw (q2) edge[sloped, anchor=center, bend left, below] node{$a, \left[4, +\right), \bot$} (q0);
			\draw (q1) edge[loop left, line width=0.4mm] node{$a, \left[0, +\right), \bot$} (q1);
			\draw (q0) edge[sloped, above, bend right=50] node{$a, \left(9, \infty\right), \top$} (q1);
			\node [below= 0cm of $(q1)!0.5!(q2)$] {$\mathcal{H}_7$};
		\end{tikzpicture}
	}	
\end{minipage}
\subcaption{Step 7}
\end{subfigure}

\vspace*{-0.3cm}
\caption{The detailed learning steps for the example DOTA $\mathcal{A}$ in Fig.~\ref{fig:COTA}, (Continued).}
\label{fig:detailed_steps}
\end{figure}

\clearpage

\section{Timed Mealy machines}
Here, we present an example of timed Mealy machine and some modification of our learning algorithm on TMMs.

\subsection{An example of TMM}
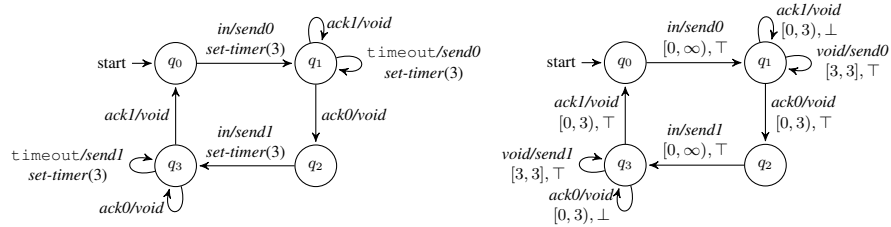
\begin{figure}[h]
\centering
\begin{minipage}{0.49\textwidth}
\centering
\vspace{0.4cm}
\begin{tikzpicture}[->, >=stealth', shorten >=1pt, auto, node distance=2cm, scale = 0.7,every node/.style={scale=0.7}]
	\node[initial, state]  (0) {$q_0$};
	\node[state](1) [right=1.3cm of 0] {$q_1$};
	\node[state](2) [below=0.8cm of 1] {$q_2$};
	\node[state](3) [below=0.8cm of 0] {$q_3$};
	
	\path (0) edge node[above, align=center] {\textit{in/send0}\\\textit{set-timer}(3)} (1) 
	(1) edge [loop above] node[right, pos=.7] {\textit{ack1/void}} (1)
	(1) edge [loop right] node[right, align=center] {\texttt{timeout}\textit{/send0}\\\textit{set-timer}(3)} (1)
	(1) edge node[right] {\textit{ack0/void}} (2)
	(2) edge node[above, align=center] {\textit{in/send1}\\\textit{set-timer}(3)} (3)
	(3) edge [loop below] node[left, pos=.7] {\textit{ack0/void}} (3)
	(3) edge [loop left] node[left, align=center] {\texttt{timeout}\textit{/send1}\\\textit{set-timer}(3)} (3)
	(3) edge node[left] {\textit{ack1/void}} (0);
\end{tikzpicture}
\label{fig:MM1T}
\end{minipage}
\hfill
\begin{minipage}{0.49\textwidth}
\centering
\begin{tikzpicture}[->, >=stealth', shorten >=1pt, auto, node distance=2cm, scale = 0.7,every node/.style={scale=0.7}]
	\node[initial, state]  (0) {$q_0$};
	\node[state](1) [right=1.3cm of 0] {$q_1$};
	\node[state](2) [below=0.8cm of 1] {$q_2$};
	\node[state](3) [below=0.8cm of 0] {$q_3$};
	
	\path (0) edge node[above, align=center] {\textit{in/send0}\\$[0,\infty),\top$} (1) 
	(1) edge [loop above] node[right, pos=.7, align=center] {\textit{ack1/void}\\$[0,3),\bot$} (1)
	(1) edge [loop right] node[right, align=center] {\textit{void/send0}\\$[3,3],\top$} (1)
	(1) edge node[right, align=center] {\textit{ack0/void}\\$[0,3),\top$} (2)
	(2) edge node[above, align=center] {\textit{in/send1}\\$[0,\infty),\top$} (3)
	(3) edge [loop below] node[left, pos=.7, align=center] {\textit{ack0/void}\\$[0,3),\bot$} (3)
	(3) edge [loop left] node[left, align=center] {\textit{void/send1}\\$[3,3],\top$} (3)
	(3) edge node[left, align=center] {\textit{ack1/void}\\$[0,3),\top$} (0);
\end{tikzpicture}
\label{fig:TMM}
\end{minipage}
\caption{Left: MM1T model of alternating-bit protocol sender; Right: the corresponding TMM model. In MM1T, at any time, if it receives a input action \textit{in} and moves from $q_0$ to $q_1$, the timer is set to 3 and starts countdown in $q_1$. If receiving no acknowledgement in three time units, \texttt{timeout} is triggered and the timer is reset to 3 for waiting for \textit{ack0}. In TMM, \texttt{timeout} is replaced by a new special input action \textit{void} since a clock replaces the timer. We do not need to copy $q_1$ (also $q_3$) since the timer values set by different transitions are the same.}
\label{fig:MM1T_and_TMM}
\end{figure}

Fig.~\ref{fig:MM1T_and_TMM} illustrates TMM model and MM1T model of alternating-bit protocol sender described in~\cite{VaandragerB021}.

Compared to MM1T where the timer can be set to integer values on transitions and affect the behavior of the system through timeouts only, a timed Mealy machine is equipped with a clock making it more convenient to set timing constraints which have both upper and lower bounds. 
We can transform a Mealy machine with one timer into a corresponding timed Mealy machine. One technical issue is that we need to copy a location if it is set different timer values on different incoming transitions.  

Compared to OTAs, TMMs are more convenient for modeling reactive systems with input/output behavior. Since TMMs allow an input-output pair on each transition, we do not need to represent a pair of input/output as two transitions. Moreover, the outputs are provided by the teacher when learning TMMs, instead of having to be learned in the OTA case. This corresponds more naturally to the learning scenario for systems with input/output behavior. All these are motivations to extend our learning algorithm to DTMMs.

\subsection{Modifications of learning algorithm}
\label{sbsc:TMMmodification}

Given a timed word $\omega$, a DTMM produces an output sequence rather than whether $\omega$ is accepted. This provides richer information for distinguishing between timed words. More precisely, the membership query is modified so that on given a timed word over input actions $\omega$, the teacher returns the output sequence corresponding to $\omega$. For the equivalence query, the teacher determines whether $\mathcal{M}_{\mathcal{H}} \approx \mathcal{M}$ for the current hypothesis DTMM $\mathcal{M}_{\mathcal{H}}$.

The learning algorithm is modified according to the changes in the membership query. In particular, the procedure for distinguishing two rows under last resets (the test $T(\omega_1,\omega_2,i_1,i_2,e)$ in Section~\ref{sec:alignment-comparison}) is modified. Instead of comparing whether the two timed words are accepted, we compare the output of the two queries \emph{in the portion of suffix $e$}. This is because the output on $\omega_1$ and $\omega_2$ may be different even if they reach the same location. Hypothesis construction (Section~\ref{sec:hypothesis-construction}) is modified to take into account additional output information from queries. The other parts of the algorithm are unchanged. The modification from DOTA to DTMM case is similar to the modification from learning DFA to Mealy machines, for example described in~\cite{ShahbazG09}.

\end{subappendices}

\end{document}